\documentclass[prd,aps,floats,floatfix,eqsecnum,nofootinbib]{revtex4}
\usepackage{graphicx,amssymb,amsbsy,bm,amsmath,rotating}
\usepackage{psfrag}
\usepackage{color}
\newcommand{\be}{\begin{equation}}
\newcommand{\ee}{\end{equation}}
\newcommand{\bea}{\begin{eqnarray}}
\newcommand{\eea}{\end{eqnarray}}

\begin{document}
\title{Higher order terms in the inflaton potential
and the lower bound on the tensor to scalar ratio r}
\author{\bf C. Destri} \email{Claudio.Destri@mib.infn.it}
\affiliation{Dipartimento di Fisica G. Occhialini, Universit\`a
Milano-Bicocca and INFN, sezione di Milano-Bicocca, 
Piazza della Scienza 3, 20126 Milano, Italia.}
\author{\bf H. J. de Vega}
\email{devega@lpthe.jussieu.fr} \affiliation{LPTHE, Universit\'e
Pierre et Marie Curie (Paris VI) et Denis Diderot (Paris VII),
Laboratoire Associ\'e au CNRS UMR 7589, Tour 24, 5\`eme. \'etage, 
Boite 126, 4, Place Jussieu, 75252 Paris, Cedex 05, France}
\affiliation{Observatoire de Paris, LERMA. Laboratoire
Associ\'e au CNRS UMR 8112.
 \\61, Avenue de l'Observatoire, 75014 Paris, France.}
\author{\bf N. G. Sanchez}
\email{Norma.Sanchez@obspm.fr} \affiliation{Observatoire de Paris,
LERMA. Laboratoire Associ\'e au CNRS UMR 8112.
 \\61, Avenue de l'Observatoire, 75014 Paris, France.}
\date{\today}
\begin{abstract}
The MCMC analysis of the CMB+LSS data in the context of the Ginsburg-Landau
approach to inflation indicated that the fourth degree double--well 
inflaton potential in new inflation gives an excellent fit of the present CMB and LSS data. 
This provided a  {\bf lower bound}
for the ratio $ r $ of the tensor to scalar fluctuations and as most probable value 
$ r \simeq 0.05 $, within reach of the forthcoming CMB observations.
In this paper we systematically analyze the effects of arbitrarily {\bf higher
order} terms in the inflaton potential on the CMB observables: spectral index $ n_s $
and ratio $ r $. Furthermore, we compute in close form the inflaton potential 
dynamically generated when the inflaton field is a fermion condensate
in the inflationary universe. This inflaton potential turns out to belong to
the Ginsburg-Landau class too. The theoretical values in the $ (n_s,r) $ plane for 
all double well inflaton potentials in the Ginsburg-Landau approach
(including the potential generated by fermions) fall inside a {\bf universal} 
banana-shaped region $ \cal B $. The upper border of the banana-shaped 
region $ \cal B $ is given by the fourth order double--well potential and provides an 
upper bound for the ratio $ r $. The lower border of $ \cal B $ 
is defined by the quadratic plus an infinite barrier inflaton potential 
and provides a {\bf lower bound} for the ratio $ r $. For example, the current best 
value of the spectral index $ n_s = 0.964 $, implies  $ r $ is in the interval: 
$ 0.021 < r < 0.053$. Interestingly enough, this range is within reach of 
forthcoming CMB observations. 
\end{abstract}
\maketitle
\tableofcontents

\section{Introduction}

The current WMAP data are validating the single field slow-roll scenario 
\cite{WMAP5}. Single field slow-roll models provide an appealing, 
simple and fairly generic description of inflation \cite{libros,revius}. This 
inflationary scenario can be implemented using a scalar field, 
the \emph{inflaton} with a Lagrangian density 
\be
\mathcal{L} = a^3(t)\left[\frac{\dot{\varphi}^2}2 -
\frac{(\nabla\varphi)^2}{2 \; a^2(t)}-V(\varphi) \right] \; ,
\ee 
where $ V(\varphi) $ is the inflaton potential. Since the universe
expands exponentially fast during inflation, gradient terms 
are  exponentially suppressed and can be neglected.
At the same time, the exponential stretching of spatial lengths
classicalize the physics and permits a classical treatment.
One can therefore consider an homogeneous and classical inflaton field 
$ \varphi(t) $ which obeys the evolution equation
\be\label{eqno} 
{\ddot \varphi} + 3 \, H(t) \; {\dot \varphi} + V'(\varphi) = 0 \; ,
\ee 
in the isotropic and homogeneous Friedmann-Robertson-Walker (FRW) metric 
\be\label{FRW}
 ds^2= dt^2-a^2(t) \; d\vec{x}^2 \quad ,
\ee
which is sourced by the inflaton.
Here $ H(t) \equiv {\dot a}(t)/a(t) $ stands for the Hubble parameter.
The energy density and the pressure for a spatially homogeneous inflaton 
are given by
\be\label{enerpres} 
\rho = \frac{\dot{\varphi}^2}2+ V(\varphi)
\quad , \quad p  =\frac{\dot{\varphi}^2}2-V(\varphi) \; . 
\ee
The scale factor $ a(t) $ obeys the Friedmann equation,
\be\label{frinf}
H^2(t) = \frac1{3 M^2_{Pl}} \left[\frac12 \; \dot \varphi^2 + 
V(\varphi)\right] \; .
\ee
In order to have a finite number of inflation efolds, the inflaton 
potential $ V(\varphi) $ must vanish at its absolute minimum
\be\label{minV}
V'(\varphi_{min})=V(\varphi_{min}) = 0 \; . 
\ee
 These two conditions guarantee that inflation is not eternal.
Since the inflaton field is space-independent inflation 
is followed by a matter dominated era (see for example ref. \cite{bibl}). 

\medskip

Inflation as known today should be considered as an {\bf effective theory},
that is, it is not a fundamental theory but a theory of a
condensate (the inflaton field) which follows from a more fundamental one.
In order to describe the cosmological evolution it is enough to consider
the effective dynamics of such condensates. The inflaton field $ \phi $ may {\bf not}
correspond to any real particle (even unstable) but is just an {\bf effective}
description while the microscopic description should come from a Grand Unification
theory (GUT) model. 

At present, there is no derivation of the inflaton model from
a microscopic GUT theory.  However, the relation between
the effective field theory of inflation and the
microscopic fundamental theory is akin to the relation between the
effective Ginsburg-Landau theory of superconductivity \cite{gl} and the
microscopic BCS theory, or like the relation of the $ O(4) $ sigma
model, an effective low energy theory of pions, photons and nucleons 
(as skyrmions), with the corresponding microscopic theory: quantum chromodynamics (QCD).

\medskip

In the absence of a microscopic theory of inflation, we find that the
Ginsburg-Landau approach is a powerful effective theory description. 
Such effective approach has been fully
successful in several branches of physics when the microscopic theory
is not available or when it is very complicated to solve in the regime considered.
This is the case in statistical physics, particle
physics and condensed matter physics. Such GL effective theory approach permits to
analyse the physics in a quantitative way without committing to a specific model \cite{gl}.

\medskip

The Ginsburg-Landau framework is not just a class of physically well
motivated inflaton potentials, among them the double and single well
potentials. The Ginsburg-Landau approach provides the effective theory for
inflation, with powerful gain in the physical insight and analysis of the
data. As explained in this paper and shown in the refs. \cite{mcmc}-\cite{bibl}, 
the analysis of the present set of CMB+LSS data with the effective theory of inflation, favor
the double well potential. Of course, just analyzing the present data
without this powerful physical theory insight, does not allow
to discriminate between classes of models, and so, very superficially
and incompletely, it would seem that almost all the potentials are still
at the same footing, waiting for the new data to discriminate them.

\medskip

In the Ginsburg-Landau spirit the potential is a polynomial in 
the field starting by a constant term \cite{gl}. Linear terms can always 
be eliminated by a constant shift of the inflaton field. 
The quadratic term can have a positive or a negative sign 
associated to unbroken symmetry (chaotic inflation) or to broken symmetry
(new inflation), respectively.

As shown in refs. \cite{mcmc,bibl} a negative quadratic term 
and a negligible cubic term in new inflation provides a very good fit to the CMB+LSS data, 
(the inflaton starts at or very close to the false vacuum $ \varphi = 0 $).
The analysis in refs.\cite{mcmc,bibl} showed that chaotic 
inflation is clearly disfavoured compared with new inflation. 
Namely, inflaton potentials with $ V''(0) < 0 $  are favoured  with the inflaton starting
to evolve at  $ \varphi = 0 $. 

We can therefore ignore the linear and cubic terms in $ V(\varphi) $. I
f we restrict ourselves for the moment to 
fourth order polynomial potentials, eq.(\ref{minV}) and  $ V''(0) < 0 $ imply
that the inflaton potential is a double well (broken symmetric) with the following form:
\be\label{bini}
V(\varphi) = - \frac12 \, m^2 \; \varphi^2 + \frac14 \; \lambda \;\varphi^4 + 
\frac{m^4}{4 \, \lambda} = \frac14 \; \lambda \; 
\left(\varphi^2-\frac{m^2}{\lambda}\right)^2 \; .
\ee
The mass term $ m^2 $ 
and the coupling $ \lambda $ are naturally expressed in terms of the {\bf two} 
energy scales which are relevant in this context: the energy scale of 
inflation $ M $ and the Planck mass $ M_{Pl} = 2.43534 \; 10^{18}$ GeV,
\be
m=\frac{M^2}{M_{Pl}} \;,\quad \lambda = \frac{y}{8 \, N} \; 
\left(\frac{M}{M_{Pl}}\right)^4 \; .
\ee
Here $ y = {\cal O} (1) $ is the quartic coupling.

The MCMC analysis of the CMB+LSS
data combined with the theoretical input above yields the value 
$ y \simeq 1.26 $ for the coupling \cite{mcmc,bibl}.
$ y $ turns out to be {\bf order one} consistent with the Ginsburg-Landau 
formulation of the theory of inflation \cite{bibl}.

According to the current CMB+LSS data, this fourth order double--well potential of 
new inflation yields as most probable 
values: $ n_s \simeq 0.964 ,\; r\simeq 0.051 $ \cite{mcmc,bibl}. 
This value for $ r $ is within reach of forthcoming CMB observations \cite{apjnos}. 
For the best fit value $ y \simeq 1.26 $, the inflaton field exits the horizon 
in the negative concavity region $ V''(\varphi) < 0 $ intrinsic to new inflation.
We find for the best fit \cite{mcmc,bibl}, 
\be\label{masas}
M = 0.543 \times 10^{16} \quad {\rm GeV ~ for ~ the ~ scale ~ of  ~ 
inflation ~ and} \quad m = 1.21 \times 10^{13} \quad {\rm GeV ~ for ~ 
the ~ inflaton ~ mass.}
\ee
It must be stressed that in our approach the amplitude of scalar fluctuations
$ |{\Delta}_{k\;ad}^{\mathcal{R}}| = (4.94 \pm 0.1)\times 10^{-5} $ allows us to
completely determine the energy scale of inflation which turns out to coincide with the
Grand Unification energy scale  (well below the Planck energy scale). 
Namely, we succeed to derive the energy scale of inflation without
the knowledge of the value of $ r $ from observations. 
$ M \ll M_{Pl} $ guarantees the validity of the effective theory approach
to inflation. The fact that the inflaton mass is $ m \ll M $ implies the
appearance of infrared phenomenon as the quasi-scale invariance of the 
primordial power.

\medskip

Since the inflaton potential must be bounded from below $ V(\varphi) \geq 0 $, the
highest degree term must be even and with a positive coefficient.
Hence, we consider polynomial potentials of degree $ 2 \, n $
where $ 1 < n \leq \infty $.

The request of renormalizability restricts the degree of the inflaton 
potential to four. However, since the theory of inflation is an effective 
theory, potentials of degrees higher than four are in principle acceptable.

\medskip

A given Ginsburg-Landau potential will be 
{\bf reliable} provided it is {\bf stable} under the addition to the 
potential of terms of higher order. Namely, adding to the $ 2 \, n $th 
order potential further terms of order $ 2 \, n + 1 $ and $ 2 \, n + 2 $ should 
only produce {\bf small} changes in the observables.
Otherwise, the description obtained could not be trusted.
Since, the highest degree term must be even and positive, this
implies that all even terms of order higher or equal than four should
be positive. 

Moreover, when expressed in terms of the appropriate dimensionless variables,
a relevant dimensionless coupling constant $ g $ can be defined by rescaling
the inflaton field. This coupling $ g $ turns out to be of order $ 1/N $ where
$ N \sim 60 $ is the number of efolds since the cosmologically relevant modes 
exit the horizon till the end of inflation showing that the slow-roll approximation
is in fact an expansion in $ 1/N $ \cite{1sN}. It is then natural to introduce
as coupling constant $ y \equiv 8 \, N \; g = {\cal O} (N^0) $. This
is consistent with the stability of the results in the above sense.
Generally speaking, the Ginsburg-Landau approach makes sense for small or moderate
coupling. 

\medskip

Odd terms in the inflaton field $ \varphi $ are allowed in $ V(\varphi) $ 
in the effective theory of inflation.
Choosing $ V(\varphi) $ an even function of $ \varphi $
implies that $ \varphi \to -\varphi $ is a symmetry of the 
inflaton potential. At the moment, as stated in \cite{ciri,bibl},
we do not see reasons based on fundamental physics to choose a zero or a 
nonzero cubic term, which is the first non-trivial odd term.
Only the phenomenology, that is the fit to the CMB+LSS data, decides 
on the value of the cubic and the higher order odd terms. 
The MCMC analysis of the WMAP plus LSS data shows that the cubic term is 
negligible and therefore can be ignored for new inflation 
\cite{mcmc,bibl}. CMB data have also been analyzed at the light of 
slow-roll inflation in refs. \cite{otros}.

\medskip

In the present paper we systematically study the effects produced by
higher order terms ($ n > 4 $) in the inflationary potential on the 
observables $ n_s $ and $ r $. 

\medskip

We show in this paper that all $ r=r(n_s) $ curves for a large class of double--well
potentials of arbitrary high order in new inflation 
fall {\bf inside} the {\bf universal} banana region 
$ \cal B $ depicted in fig. \ref{banana}.
Moreover, we find that the $ r=r(n_s) $ curves for even double--well potentials 
with arbitrarily positive higher order terms lie {\bf inside} the 
universal banana region $ \cal B $ [fig. \ref{banana}]. This is true for 
arbitrarily large values of the coefficients in the potential. 

\medskip

Furthermore, the inflaton field may be a condensate of fermion-antifermion pairs 
in a grand unified theory (GUT) in the inflationary background.
In this paper we explicitly write down in closed form the inflaton potential dynamically generated
as the effective potential of fermions in the inflationary universe.
This inflaton potential turns out to belong to the Ginsburg-Landau class
of potentials considered in this paper.
We find that the corresponding $ r=r(n_s) $ curves lie inside the {\bf universal} 
banana region $ \cal B $ provided the one-loop part of the inflaton potential is
at most of the same order as the tree level piece.
Therefore, a {\bf lower bound} for the ratio tensor/scalar fluctuations $ r $ is present 
for {\bf all} potentials above mentioned. For the current best value of the spectral index 
$ n_s = 0.964 $ \cite{bibl,WMAP5} the lower bound turns out to be $ r > 0.021 $.
 
Namely, the shape of the banana region  fig. \ref{banana} {\bf combined} with the
value $ n_s = 0.964 $ for the spectral index yields the lower bound $ r > 0.021 $.
If one consider low enough values for $ n_s $ (in disagreement with observations)
$ r $ can be arbitrarily small within the GL class of inflaton potentials. 

The upper border of the universal region $ \cal B $ tells us that
$ r < 0.053 $ for $ n_s = 0.964 $.
Therefore, we have inside the region $ \cal B $ within the large class of potentials 
considered here 
$$
 0.021 < r < 0.053 \quad {\rm for}\quad  n_s = 0.964 \; .
$$
Interestingly enough $ r \simeq 0.04 $ is within reach, although borderline
for the Planck satellite \cite{apjnos}. 

\medskip

Among the simplest potentials in the Ginsburg-Landau class, the one that best reproduces the 
present CMB+LSS data, is the fourth order 
double--well potential eq.(\ref{bini}), yielding as most probable values: 
$ n_s \simeq 0.964 , \; r\simeq 0.051 $. Our work here shows that adding 
higher order terms to the inflaton potential does not really improve the 
data description in spite of the addition of new free parameters.
Therefore, the fourth order double--well potential gives a robust
and stable description of the present CMB/LSS data and provides clear predictions
to be contrasted with the forthcoming CMB observations \cite{apjnos}.

There is an abundant literature on slow-roll inflationary potentials and 
the cosmological parameters $ n_s $ and $ r $ including new inflation and 
in particular hilltop inflation \cite{lily,otros2,otros3,otros4}.

The question on whether a lower bound for $ r $ is found or
not depends on whether the Ginsburg-Landau (G-L) effective field 
approach to inflation is used or not. Namely, within  the G-L approach,
the new inflation double well potential determines
a banana shaped relationship $ r = r (n_s) $ which  for the
observed  $ n_s $ value determines a lower bound on $ r $.
The analysis of the CMB+LSS data within the G-L approach
which we performed in refs. \cite{mcmc,bibl} shows that new inflation 
is preferred by the data with respect to chaotic inflation 
for fourth degree potentials, and that the lower bound on $ r $ is then present.
Without using the powerful physical G-L framework such discrimination between 
the two classes of inflation models is not possible and the lower bound for $ r $
does not emerge. Other references in the field (i. e. \cite{lily,kinney08,PeirisEasther})  
do not work within the Ginsburg-Landau framework, do not find lower bounds for $ r $ 
and cannot exclude arbitrarily small values for $ r $, much smaller than our lower 
bound $ r \simeq 0.021 $.

\bigskip

This paper is organized as follows: in section II we present in general inflaton
potentials of arbitrary high degree, specializing then to fourth and sixth--order
polynomial potentials and displaying their corresponding $ r=r(n_s) $ curves. 
Sec. III contains the $ 2 \, n$th order double--well polynomial inflaton potentials with
arbitrary random coefficients and their $ r=r(n_s) $ curves. Sec. IV presents
the $ n \to \infty $ limits of these polynomial potentials and we present in sec. V
the exponential potential and its infinite coupling limit. In sec. V we compute
the inflaton potential from dynamically generated fermion condensates in a de Sitter 
space-time displaying their $ r=r(n_s) $ curves. Finally, we present and discuss the 
universal banana region in sec. VII together with our conclusions.

\section{Physical parametrization for inflaton potentials}\label{sec2}

We start by writing the inflaton potential in dimensionless variables as
\cite{ciri}
\be\label{eq:M}
V(\varphi) = M^4 \; v\left(\frac{\varphi}{M_{Pl}}\right)  \; ,
\ee
where $ M $ is the energy scale of inflation and $ v(\phi) $ is a 
dimensionless function of the dimensionless field argument 
$ \phi = \varphi/M_{Pl} $. 
Without loss of generality we can set $ v'(0)=0 $. Moreover, 
provided $ V''(0)\neq 0 $ 
we can choose without loss of generality $ |v''(0)|=1/2 $.

In the slow-roll regime, higher time derivatives in the equations of motion 
can be neglected with the final well known result for the number of efolds
\be\label{eq:N}
N = -\int_{\phi_{exit}}^{\phi_{end}} d\phi \; \frac{v(\phi)}{v'(\phi)} \; ,
\ee
where $ \phi_{exit} $ is the inflaton field at horizon exit. 
To leading order in $ 1/N $ we can take $ \phi_{end} $ to
be the value $ \phi_{min} $ at which $ v(\phi) $ attains its absolute 
minimum $ v(\phi_{min}) $, which must be zero since inflation must stop after 
a finite number of efolds \cite{bibl}. 

Then, in chaotic inflation we have $ \phi_{min}=0 $, with $ v'(\phi)>0 $ 
for $ \phi>0 $, while in new inflation we have $ \phi_{min}>0 $ with 
$ v'(\phi)<0 $ for $ 0<\phi<\phi_{min} $. We consider potentials 
$ v(\phi) $ that can be expanded in Taylor series around 
$ \phi=\phi_{min} $, with a non-vanishing quadratic (mass) term. 

It is convenient to rescale the inflaton field in order to
conveniently parametrize the higher order potential.
We define a coupling parameter $ g>0 $ by rescaling the inflaton and its 
potential keeping invariant the quadratic term, that is
\be \label{eq:vg}
  v(\phi)  = \frac1{g} \; v_1\left(\phi \; \sqrt{g}\right)
\ee
For a potential $ v_1(u) $ expanded in power series around $ u=0 $ we write:
\be\label{v1g}
  v_1(u) = c_0 \mp \frac12 \; u^2 +  \sum_{k \ge 3}\frac{c_{k}}{k} \, 
u^k 
\ee
Then, replacing 
\be\label{defu}
u = \phi \; \sqrt{g} \; ,
\ee
we find
\be
 v(\phi) = \frac{c_0}g \mp \frac12  \; \phi^2 + 
   \sum_{k\ge 3}\frac{g^{k/2-1}}{k} \; c_k \; \phi^k \label{vpol} \; .
\ee
The positive sign in the quadratic term corresponds to chaotic inflation (in
which case $ c_0=0 $), while the negative sign corresponds to new inflation 
(in which case $ c_0 $ is chosen such that $ v_1(u) $ vanishes at its
 absolute minimum). 

Clearly $ g $ plus the set of coefficients $ c_k $ provide an overcomplete 
parametrization of the inflaton potential which we will now reduce.
In the case of chaotic inflation a convenient choice is $ c_4=1 $, so that
\be\label{vpol1}
   v(\phi) = \frac12  \; \phi^2 + 
\sqrt{g} \; \frac{c_3}3 \; \phi^3 + \frac{g}4 \; \phi^4 +
\sum_{k\ge 5}\frac{g^{k/2-1}}{k} \; c_k \; \phi^k 
\quad {\rm [chaotic ~inflation]} 
\ee
which represents a generic higher order perturbation of the trinomial
chaotic inflation studied in refs.~\cite{mcmc}.

In the case of new inflation, where $ \phi_{min}>0 $, it is more convenient 
to set without loss of generality that  $ u_{min} = 1 , \; \phi_{min} = 
1/\sqrt{g} $.
In order to have appropriate inflation, $ u_{min} = 1 $ must be the 
absolute minimum of $ v_1(u) $ and the closest one to the origin on the 
positive semi--axis. That is,
\be\label{vinc}
 v_1'(1) = -1 + \sum_{k \ge 3} c_k = 0
\ee
and then  $ v_1(1) = 0 $ fixes from eq.(\ref{eq:vg}) the constant term 
$ c_0 $ in the potential
\be\label{c0e}
c_0 = \frac12 - \sum_{k \ge 3}\frac{c_{k}}{k}
\ee
We thus get for the inflaton potential
\be\label{v1gen}
   v_1(u) = \frac12(1-u^2) + \sum_{k \ge 3}\frac{c_{k}}{k} \; (u^k-1)
\quad {\rm [new ~inflation]} \; ,
\ee
corresponding to
\be\label{vpol2}
v(\phi) = \frac12  \left(\frac1{g} - \phi^2 \right) + 
\sum_{k\ge 3}\frac{c_k}{k} \left(g^{k/2-1} \; \phi^k - 
\frac1{g}\right) \quad {\rm [new ~inflation]} 
\ee
For the coupling $ g $ and the field $ \phi $ using eq.(\ref{defu}),
\be\label{defg}
 g = \frac1{\phi_{min}^2} = \frac{M_{Pl}^2}{\varphi_{min}^2} 
\quad , \quad u = \frac{\phi}{\phi_{min}} = 
\frac{\varphi}{\varphi_{min}} \;.
\ee
From eq.(\ref{eq:N}) it now follows that the parameter $ g $ can
be expressed as the integral 
\be \label{eq:gN}
y(u) = 8 \; \int_{u_{min}}^u dx \;  \frac{v_1(x)}{v_1'(x)}
\; , \quad u \equiv \sqrt{g} \; \phi_{exit}  \quad  , 
\ee
where,
\be \label{gy}
g = \frac{y(u)}{8\,N} \; ,
\ee
with $ u_{min}=0 $ for chaotic inflation and $ u_{min}=1 $ for new inflation.
Eq.(\ref{eq:gN}) can be regarded as a parametrization of $ g $ and $ y(u) $ 
in terms of the rescaled exit field $ u $. Clearly, {\em as a function of  
$ u $}, $ g $ is uniformly of order $ 1/N $. $ g $ is numerically of order 
$ 1/N $ as long as $ y(u) $ is of order one. As we shall see, typically
both $ u $ at horizon exit and  $ y(u) $ are of order one. We have 
$ 0 < u < 1 $ for new inflation and $ 0 < u < +\infty $ for chaotic 
inflation.

In what follows we therefore use $ y(u) $ instead of $ g $ as a 
{\bf coupling constant} and make contact with eq.(\ref{eq:M}) by setting
\be \label{wvx}
\varphi = M_{Pl} \; \sqrt{\frac{8\,N}{y}} \; u \quad , \quad
 V(\varphi) = \frac{8\,N \; M^4}{y} \; 
v_1\left(\sqrt{\frac{y}{8\,N}} \; \frac{\varphi}{M_{Pl}}\right) \; .
\ee
We can easily read from this equation the order of magnitude of 
$ \varphi $ and $ V(\varphi) $ since $ N \sim 60 , \;  M $ is given by 
eq.(\ref{masas}) and $ u $ and $ y $ are of order one. Hence,
$ \varphi \sim M_{Pl} $ and $ V(\varphi)\sim N \; M^4 $.

As we will see below, the coupling $ y $ (or $ g $) is the most relevant 
coupling since it is related to the inflaton rescaling: the
tensor--scalar ratio $ r $ and the spectral index $ n_s $ vary in a more
relevant manner with $ y $ than with the rest of the parameters 
$ c_k, \; k \ge 3 $ in the potential eq.(\ref{vpol}).

\medskip

By construction the function $ y(u) $ has the following properties
\begin{itemize}
\item $ y(u) > 0\, $; 
\item $ y'(u) > 0 $ for $ u>0 $ in chaotic inflation;
\item $ y'(u) < 0 $ for $ 0 < u < u_{min}=1 $ in new inflation;
\item 
$ y(u)  = 2 \; (u-u_{min})^2 + {\cal O}(u-u_{min})^3 \to 0 $
as $ u \to  u_{min} $;
\item $ y(u) \to \infty $ as $ u \to \infty $ in chaotic inflation;
\item $ y(u) \simeq -8 \; v_1(0) \; \log u \to +\infty $ as 
$ u \to 0^+ $ in new inflation.
\end{itemize}

In terms of this parametrization and to leading order in $ 1/N $, the 
tensor to scalar ratio $ r $ and the spectral index $ n_s $ read:
\be\label{eq:nsr}  
r = \frac{y(u)}{N} \; \left[\frac{v_1'(u)}{v_1(u)} \right]^2 \;,\quad
n_s - 1 = -\frac3{8} \; r +  \frac{y(u)}{4N} \; \frac{v_1''(u)}{v_1(u)}
\ee 
Notice that both $ n_s - 1 $ and $ r $ are of order
$ 1/N $ for generic inflation potentials in this Ginsburg-Landau framework
as we see from eq.(\ref{eq:nsr}). Moreover, the running of the scalar spectral index 
from eq.(\ref{wvx}) and its slow-roll expression turn out to be 
of order $ 1/N^2  \sim 1/3600 \sim 3 \times 10^{-4} \ll 1 $ 
$$
\frac{d n_s}{d \ln k}= - \frac{y^2(u)}{32 \, N^2} \left\{
\frac{v_1'(u) \;
v_1'''(u)}{v_1^2(u)} + 3 \; \frac{[v_1'(u)]^4}{v_1^4(u)}
-4 \; \frac{[v_1'(u)]^2 \; v_1''(u)}{v_1^3(u)}\right\} \quad .
$$
and therefore can be neglected \cite{bibl}.
Such small estimate for $ dn_s/d \log k $ is in agreement
with the present data \cite{WMAP5} and makes the running unobservable for a foreseeable future.

Since $ y=y(u) $ can be inverted for any $ 0<u<u_{min} $, these two 
relations can also be regarded as parametrizations $ r=r(y) $ and 
$ n_s=n_s(y) $ in terms of the coupling constant $ y $. 

We are interested in the region of the $ ( n_s,r) $ plane 
obtained from eq.(\ref{eq:nsr}) by varying $ y $ (or $ u $) and the other
parameters in the inflaton potential. We call $ {\cal B} $ this region.

From now on, we will restrict to new inflation.

For a generic $ v_1(u) $ [with the required global properties described
above] we can determine the asymptotic of $ {\cal B} $, since they follow
from the weak coupling limit $ y \to 0 $ and from the strong coupling limit 
$ y \to \infty $. When $ y \to 0 $, then $ u \to  u_{min}=1 $ and
from the property above,
\be\label{eq:rweak}
  r = \frac8{N} + {\cal O}(u-1)=0.13333 \ldots+ {\cal O}(u-1)
\ee
and
\be\label{eq:nsweak}
  n_s = 1 - \frac2{N} + {\cal O}(u-1)=0.9666\ldots+ {\cal O}(u-1) \; .
\ee
When $ y\to\infty $ we have in new inflation $ u\to0 $ and then,
\be \label{eq:strong}
n_s \simeq 1 + \frac2{N} \; \log u \longrightarrow -\infty \quad ,\quad 
r \simeq -\frac8{N} \; \frac{u^2 \; \log u}{v_1(0)} \longrightarrow 0^+ \; .
\ee
We see that in the strong coupling regime $ r $ becomes very small and 
$ n_s $ becomes well below unity. However, the slow-roll approximation is 
valid for $ |n_s - 1| < 1 $ and in any case, the WMAP+LSS results exclude 
$ n_s < 0.9 $ \cite{WMAP5}. Therefore, the strong coupling limit is ruled 
out. 

\medskip

Eq.(\ref{eq:nsr}) for $ r $ can be rewritten using eq.(\ref{eq:gN}) 
in the suggestive form,
\be \label{rygran}
r = \frac{64}{N \; y(u)} \; \left[\frac{d \, \ln y(u)}{du}\right]^{-2}
\ee
Since $ 64/N \sim 1 , \; r $ may be {\bf small} only in case $ y(u) $ is 
{\bf large} (the logarithmic derivative of $ y(u) $ has a milder effect 
for large $ y(u) $.)
Therefore, we only find $ r \ll 1 $ in a {\bf strong} coupling regime.
Notice that $ \varphi $ is much smaller than $ M_{Pl} $ in the strong
coupling regime [eq.(\ref{defg})].

\medskip

Let us now study large classes of physically meaningful inflaton potentials
in order to provide generic bounds on the region $ {\cal B} $ of the  
$ (n_s,r) $ plane within an interval of $ n_s $ surely compatible with the 
WMAP+LSS data for  $ n_s $, namely
 $ 0.93 < n_s < 0.99 $. To gain insight into the problem,
we consider first the cases amenable to an analytic
treatment, leaving the generic cases to a numerical investigation.
As we will see below, the boundaries of the region $ {\cal B} $ turn out to be
described parametrically by the analytic formulas 
(\ref{binsr}) and (\ref{nsrinf}). 

\subsection{The fourth degree double--well inflaton potential}\label{sec:dbw}

The case when the $ V(\varphi) $ is the standard double--well quartic 
polynomial 
\begin{equation*}
V(\varphi) = \frac14 \; \lambda \; 
\left(\varphi^2-\frac{m^2}{\lambda}\right)^2
\end{equation*}
has been studied in refs. \cite{mcmc,bibl}.
In the general framework outlined above we have for this case,
\be\label{binon}
v_1(u) = \frac14 \; (u^2-1)^2 = \frac14 - \frac12 \; u^2 + \frac14 \; u^4
\; , \quad   \lambda = \frac{y}{8 \, N} \; 
\left(\frac{M}{M_{Pl}}\right)^4 \; , \quad m=\frac{M^2}{M_{Pl}} \; .
\ee
By explicitly evaluating the integral in eq.~(\ref{eq:gN}) one obtains
\be\label{eq:yexpl4}
  y(u) = u^2 - 1 -\log u^2 \; , 
\ee
and then, from eq.~(\ref{eq:nsr})
\be\label{binsr}
n_s = 1 - \frac1{N} \; \frac{3 \; u^2  + 1}{(1-u^2)^2}\, (u^2 - 1 -\log u^2)
 \;,\quad r = \frac1{N} \;  \frac{16\,u^2}{(1-u^2)^2}\,(u^2 - 1 -\log u^2)
\ee
where $ 0 \leq u \leq u_{min} = 1 $. As required by the general arguments 
above, $ u $ is a monotonically decreasing function of $ y $, ranging from 
$ u = 1 $ till $ u = 0 $ when $ y $ increases from $ y = 0 $ till 
$ y = +\infty $. In particular, when $ u \to 1^- , \; y $
vanishes quadratically as,
$$
y(u) \buildrel{u \to 1^-}\over= \frac12 \; (1-u^2)^2 \; .
$$
The concavity of the potential eq.(\ref{binon}) for the inflaton field 
at horizon crossing takes the value
$$
v_1''(u) = 3 \; u^2  - 1 \; .
$$
We see that $ v_1''(u) $ vanishes at $ u = 1/\sqrt3 $, that is at
$ y = \ln 3 - 2/3 = 0.431946\ldots $. (This is usually called
the spinodal point \cite{revi}). Therefore,
\be\label{sigwseg}
v_1''(u) > 0 \quad {\rm for}  \quad y < 0.431946\ldots
\quad {\rm and} \quad v_1''(u) < 0 \quad {\rm for}  \quad y > 
0.431946\ldots \; .
\ee
Our MCMC analysis of the CMB+LSS data combined with the theoretical
model eq.(\ref{binon}) yields $ y \simeq 1.26 $ \cite{mcmc,bibl}
deep in the {\bf negative} concavity region $ v_1''(u) < 0 $.

The negative concavity case  $ v_1''(u) < 0 $ for
$ y > 0.431946\ldots $ is {\bf specific} to new inflation 
eq.(\ref{binon}). $ v_1''(u) $ can be expressed as a linear
combination of the observables $ n_s $ and $ r $ as 
$$
n_s - 1 + \frac38 \; r = \frac{y(u)}{4N} \; \frac{v_1''(u)}{v_1(u)}
$$
As expected in the general framework presented above, the limit
$ u \to 1^- $ implies weak coupling
$ y \to 0^+ $, that is, the potential is quadratic around the absolute
minimum $ u_{min} = 1 $ and we find,
\be\label{cobino}
 n_s \buildrel{y \to 0}\over= 1 - \frac2{N} 
\quad , \quad  r \buildrel{y \to 0}\over= \frac8{N} \quad , \quad
u  \buildrel{y \to 0}\over= 1 \; , 
\ee
which coincide with $ n_s $ and $ r $ 
for the monomial quadratic potential in chaotic inflation.

In the limit $ u \to 0^+ $ which implies $ y\to +\infty $ (strong 
coupling), we have
$$ 
u \buildrel{y \to +\infty}\over=  e^{-(y+1)/2} \to 0^+ 
$$
and
\be\label{binonewyG}
n_s \; \; \buildrel{y \gg 1}\over=1 - \frac{y}{N}
\quad , \quad r \; \;  \buildrel{y \gg 1}\over= \frac{16 \; y}{N} \; 
e^{-y-1} \;  .
\ee
Notice that the slow-roll approximation is no longer valid when the 
coefficient of $ 1/N $ becomes much larger than unity. Hence, the results 
in eq.(\ref{binonewyG}) are valid for $ y \lesssim N $. We see that in 
this strong coupling regime (see fig. \ref{ban6}), $ r $ becomes very 
small and $ n_s $ becomes well below unity. However, the WMAP+LSS results 
exclude $ n_s \lesssim 0.9 $ \cite{WMAP5}. Therefore, this strong coupling 
limit $ y \gg 1 $ is ruled out. 

\medskip

For the fourth order double--well inflaton potential, the relation 
$ r = r(n_s) $ defined by eq.(\ref{binsr}) is a single curve depicted with 
dotted lines in fig. \ref{ban6}. It represents the upper border of the 
banana shaped region $ \cal B $ in fig. \ref{ban6}.

Notice that there is here a maximum value for $ n_s $, namely  
$ n_s^{max} =  
0.96782 \ldots $ with $ r(n_s^{max}) =  0.1192\ldots$ \cite{bibl}. 
The curve $ r = r(n_s) $ has here {\bf two branches}: the lower branch 
$ r < r(n_s^{max}) $ in which $ r $ {\bf increases} with increasing 
$ n_s $ and the upper branch  $ r > r(n_s^{max}) $ 
in which $ r $ {\bf decreases} with increasing $ n_s $.

\subsection{The sixth--order double--well inflaton potential}\label{sextico}

We consider here new inflation described by a six degree even polynomial
potential with broken symmetry. According to eq.~(\ref{eq:M}) and
eq.~(\ref{eq:vg}) we then have
\be\label{v6f}
V(\varphi) = \frac{M^4}g \;v_1\left(\frac{\sqrt{g} \; \varphi}{M_{Pl}}\right)
\;, \quad v_1(u) = c_0 -\frac12 \; u^2 + \frac{c_4}4 \; u^4 + 
\frac{c_6}6 \; u^6 \; .
\ee
where for stability we assume $ c_6\ge0 $. Moreover, if we regard this case 
as a higher order correction to the quartic double--well potential, then 
$ c_4 $ is positive. 

The inflaton potential eq.(\ref{v6f}) is a particular case of eq.(\ref{v1g}).
The conditions eqs. (\ref{vinc}) and (\ref{c0e}) that the absolute minimum of $ v_1(u) $
be at $ u_{min} = 1 $ yields
\be
c_4+c_6=1 \quad , \quad c_0 = \frac12 - \frac14\,c_4 - \frac16\,c_6
\ee
It is convenient to use $ b\equiv c_6 $ as free parameter so that 
$ b \geq 0 $ and $ c_4=1-b $. Thus, 
\be\label{v1s}
v_1(u) =  \frac12\;(1-u^2) -\frac{1-b}4 \; (1-u^4) - \frac{b}6 \; (1-u^6) = 
  \frac1{12} \; (1-u^2)^2 \; (3+b+2\,b \; u^2)
\ee
where $ b \leq 1 $ in order to ensure that $ c_4 \geq 0 $.

The integral in eq.~(\ref{eq:gN}) can be explicitly evaluated with the result
\be \label{eq:yexpl6}
    y(u) = 8\int_1^u dx \; \frac{v_1(x)}{v_1'(x)} 
    = \frac23 \; (u^2-1) - \frac13 \; (3+b) \; \log u^2 
    + \frac{(1+b)^2}{3\,b} \; \log\frac{1+b\,u^2}{1+b}
\end{equation}
According to the general arguments presented above [see the
lines below eq.~(\ref{eq:gN})] one can verify that  $ y(u) $ is a 
monotonically decreasing function of $ u $ for $ 0<u<1 $, where 
$ +\infty>y>0 $.

\medskip

\begin{figure}[ht]
\includegraphics[height=9cm]{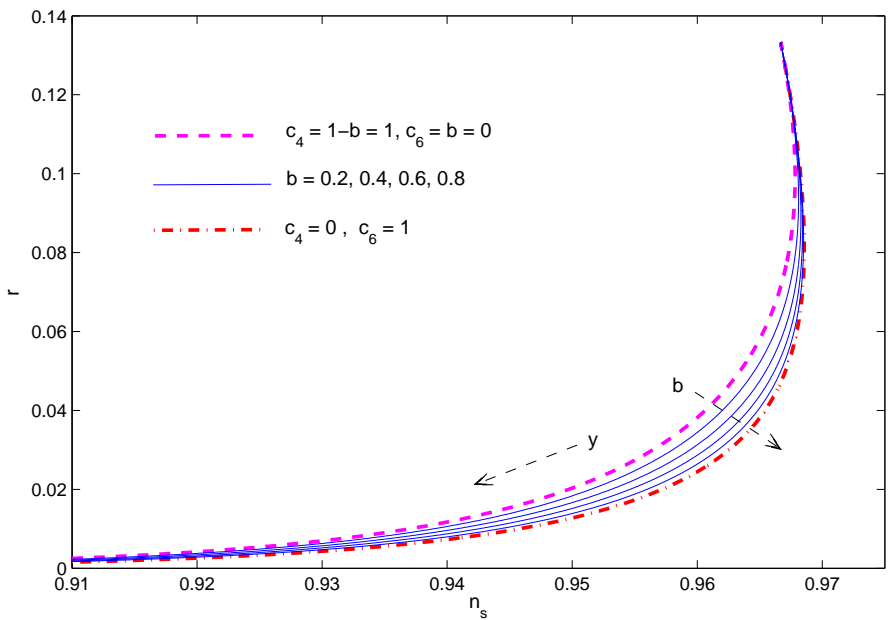}
\caption{We plot here $ r $ vs. $ n_s $ for the broken--symmetry sixth--order
inflaton potential  eq.(\ref{v1s}) setting $ N = 60 $.  The curves are obtained from
eq.~(\ref{eq:nsr6}) with the sextic coefficient $ b \equiv c_6 $ fixed to the values 
indicated in the figure. We see that $ y $ is the relevant coupling while $ b $ only 
varies $ r $ and $ n_s $ transversely to the narrow banana-shape region. The two
important limiting curves are shown: $ b = 0 $ corresponding to the fourth degree
potential eq.(\ref{binon}) and  $ b = 1 $ corresponding to the sixth degree 
potential eq.(\ref{buno}).
The uppermost point where all curves coalesce corresponds to the monomial quadratic
potential $ n_s = 0.9666\ldots, \; r = 0.13333 \ldots $ for $ N=60 $ [see
eqs.(\ref{eq:rweak})-(\ref{eq:nsweak})].}
\label{ban6}
\end{figure}

The scalar index $ n_s $ and the tensor--scalar ratio $ r $ are evaluated 
from eq.~(\ref{eq:nsr}) as
\be\label{eq:nsr6}
  r = \frac{y}{N} \; \left[\frac{12\,u \; (1+b\,u^2)}{(1-u^2)\,
      (3+b+2\,b \; u^2)}\right]^2 \; , \quad
n_s = 1 - \frac3{8} \; r +\frac{3\,y(u)}{N} \; \frac{5\,b \; u^4+ 3\,(1-b) \; u^2-1}
{(1-u^2)^2 \; (3+b+2\,b \; u^2)}
\ee
Various curves $ r=r(n_s) $ are plotted in fig.~\ref{ban6} for several values of
$ b $ in the interval $ [0,1] $ sweeping the region $ \cal B $.  We see that for 
increasing $ b $ [namely, for increasing sextic coupling and decreasing quartic coupling, 
see eq.(\ref{v1s})] the curves move down and right, sweeping the banana-shape region 
$ \cal B $ depicted on fig.~\ref{ban6}.

Clearly, $ y $ is a  variable more {\bf relevant} than $ b $. Changing $ y $
moves $ n_s $ and $ r $ in the whole available range of values,
while changing $ b $ only amounts to displacements {\bf transverse} to the
banana region $ \cal B $ in the $ n_s , \; r $ plane. In particular, for a given 
$ n_s , \; r $ becomes smaller for increasing $ b $.

\medskip

We see in fig.~\ref{ban6} two important {\bf limiting} curves: the $ b\to 0 $ and the $
b \to 1 $ curves. When $ b=0 $ the function $ v_1(u) $ reduces to the fourth order
double-well potential eq.(\ref{binon}) and we recover its
characteristic curve $ r=r(n_s) $. When $ b=1 $ the potential has no quartic term
and reduces to the quadratic plus sixth order potential:
\be\label{buno}
v_1(u) \buildrel{b \to 1}\over= \frac16\,(1-u^2)^2 \; (2+u^2) = 
\frac13 -\frac12\,u^2 + \frac16\,u^6 \; . 
\ee
In summary, the quadratic plus quartic broken--symmetry potential describes the
upper/left border of the banana--shaped region $ {\cal B} $ of fig.~\ref{ban6}, 
while the quadratic plus sextic broken--symmetry
potential describes its lower/right border.

\section{Higher--order even polynomial double-well inflaton 
potentials}\label{pares}

The generalization of the sixth order inflaton potential with broken 
symmetry to arbitrarily higher orders is now straightforward:
\be\label{eq:vg2n}
  V(\varphi) = \frac{M^4}g \; v_1\left(\frac{\sqrt{g} \; \varphi}{M_{Pl}}\right)
  \;,\quad v_1(u) = \frac12\,(1-u^2) + \sum_{k=2}^{n}\frac{c_{2k}}{2k} \; (u^{2k}-1)\; , 
\ee
with the constraint eq.(\ref{vinc}) 
\be\label{eq:norm}
  \sum_{k=2}^{n}\,c_{2k} = 1
\ee
which guarantees that $ u=1 $ is an extreme of $ v_1(u) $. 

We consider here the case when all higher coefficients $ c_{2k} $ are 
positive or zero :
\begin{equation*}
  c_{2k} \ge 0 \;,\quad k = 2,\ldots,n
\end{equation*}
such that  $ u_{min}=1 $ is the unique positive minimum.

\begin{figure}[ht]
\includegraphics[height=10cm]{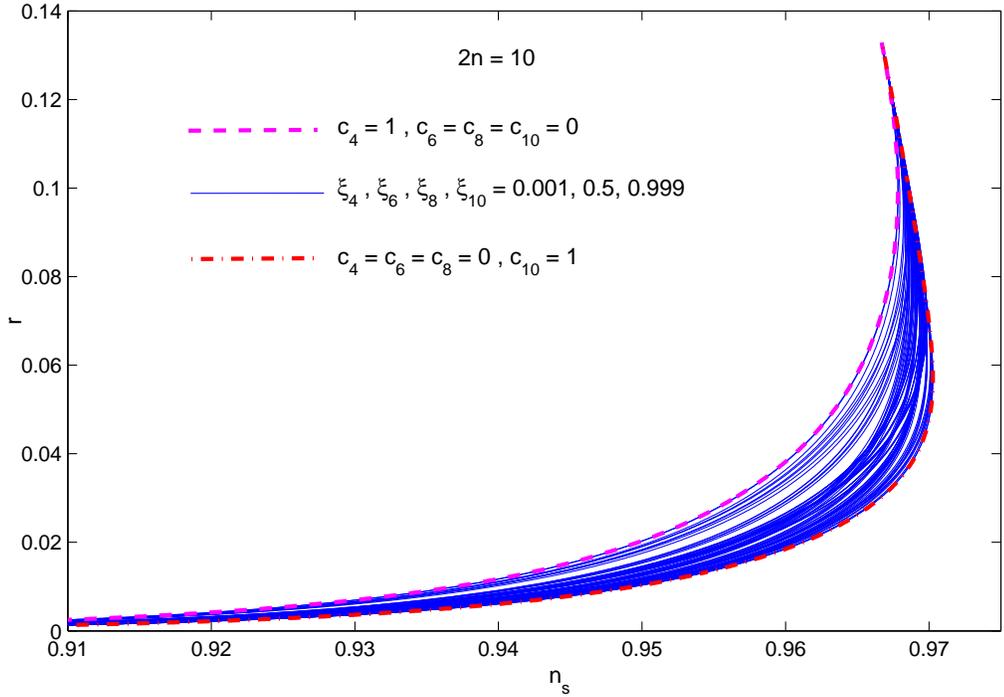}
\caption{$ r $ vs. $ n_s $ for the $10$th. order even polynomial potential
eq.~(\ref{eq:vg2n}) with $ n=5 $ and the coefficients $ c_{2k} $ taking
independently the values indicated. The relation to the numbers $ \xi_k $ is
given in eq.~(\ref{eq:xi2c}). The upper/left border curve $ c_4=1 , \; 
c_6=c_8=c_{10} = 0 $ corresponds to the fourth order potential 
eq.(\ref{binon}). The lower/right border curve $ c_{10}=1 , \;  c_4=c_6=c_8 
= 0 $ corresponds to the quadratic plus 10th order
term potential eq.(\ref{wn}) for $ n = 5 $. 
These are the limiting curves of the banana $ {\cal B} $ region.}
\label{fig:ban10}
\end{figure}

\begin{figure}[h]
\includegraphics[height=10cm]{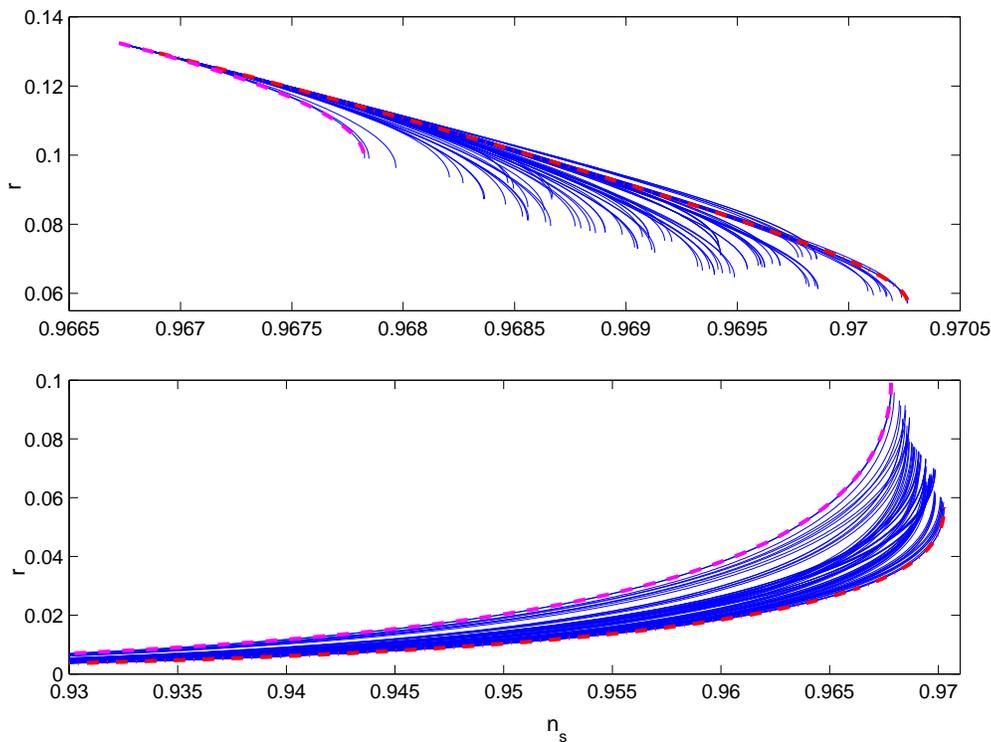}
\caption{A detail 
The banana region $ \cal B $ in the $ (n_s,r) $ plane for the quadratic 
plus $10$th order polynomial as in fig.~\ref{fig:ban10}, but with the 
curves split in two parts by the value $ r(n_s^{max}) $. The upper
panel shows the upper branches  $ r > r(n_s^{max}) $ in which
$ r $ {\bf decreases} with $ n_s $ while the lower panel shows the lower 
branches  $ r < r(n_s^{max}) $ in which $ r $ {\bf increases} with 
$ n_s $. The quadratic plus $10$th order polynomial thus provides the
{\bf lower border} of the banana region $ \cal B $ setting the lower 
bound on $ r $. This bound is here $ r > $ for the observed allowed
range $ < n_s < $.}
\label{fig:splitban10}
\end{figure}

We determine the shape of the $ {\cal B} $ region
for arbitrary positive or zero values of the coefficients $ c_{2k} $
[subject to the constraint (\ref{eq:norm})], 
performing a large number of simulations with different setups. After 
producing coefficients $ c_{2k} $ we numerically computed the function 
$ y(u) $ following eq.(\ref{eq:gN})
\begin{equation*}
  y(u) = 4 \int^1_u  \frac{dx}{x} \; 
\frac{1 - x^2 + \sum_{k=2}^n \displaystyle \frac{c_{2k}}{k} \; (x^{2k}-1)}
  {1 - \sum_{k=2}^n c_{2k} \; x^{2k-2} }
\end{equation*}
and obtain the $ r=r(n_s) $ curves from eq.~(\ref{eq:nsr}) by plotting directly 
$ r $ vs. $ n_s $.

Uniform distributions of coefficients are obtained by setting
\be\label{eq:xi2c}
c_{2k} = \left(\sum_{j=1}^n \log \xi_j\right)^{-1} \; \log \xi_k \; ,\quad 
k = 1,2,\ldots,n 
\ee
where the numbers $ \xi_k $ are independently and uniformly distributed in the unit
interval. We used the parametrization eq.(\ref{eq:xi2c}) also when the $ \xi_k $
are chosen according to other rules.

For example, in figs.~\ref{fig:ban10}-\ref{fig:splitban10} we plot the results
when $ n=5 $, that is for the ten degree polynomial. In this case we let $ \xi_4 ,
\; \xi_6, \; \xi_8 $ and $ \xi_{10} $ take independently the values $ 0.001 , \; 0.5 $ or
$ 0.999 $, for a total of 78 distinct configurations of coefficients. For better
clarity, in figs.~\ref{fig:ban10}-\ref{fig:splitban10} we also include the two
border cases $ c_4=1 , \; c_6=c_8=c_{10} = 0 $ and $ c_{10}=1 , \;  c_4=c_6=c_8 = 0 $.

\medskip

\begin{figure}[h]
\includegraphics[height=10cm]{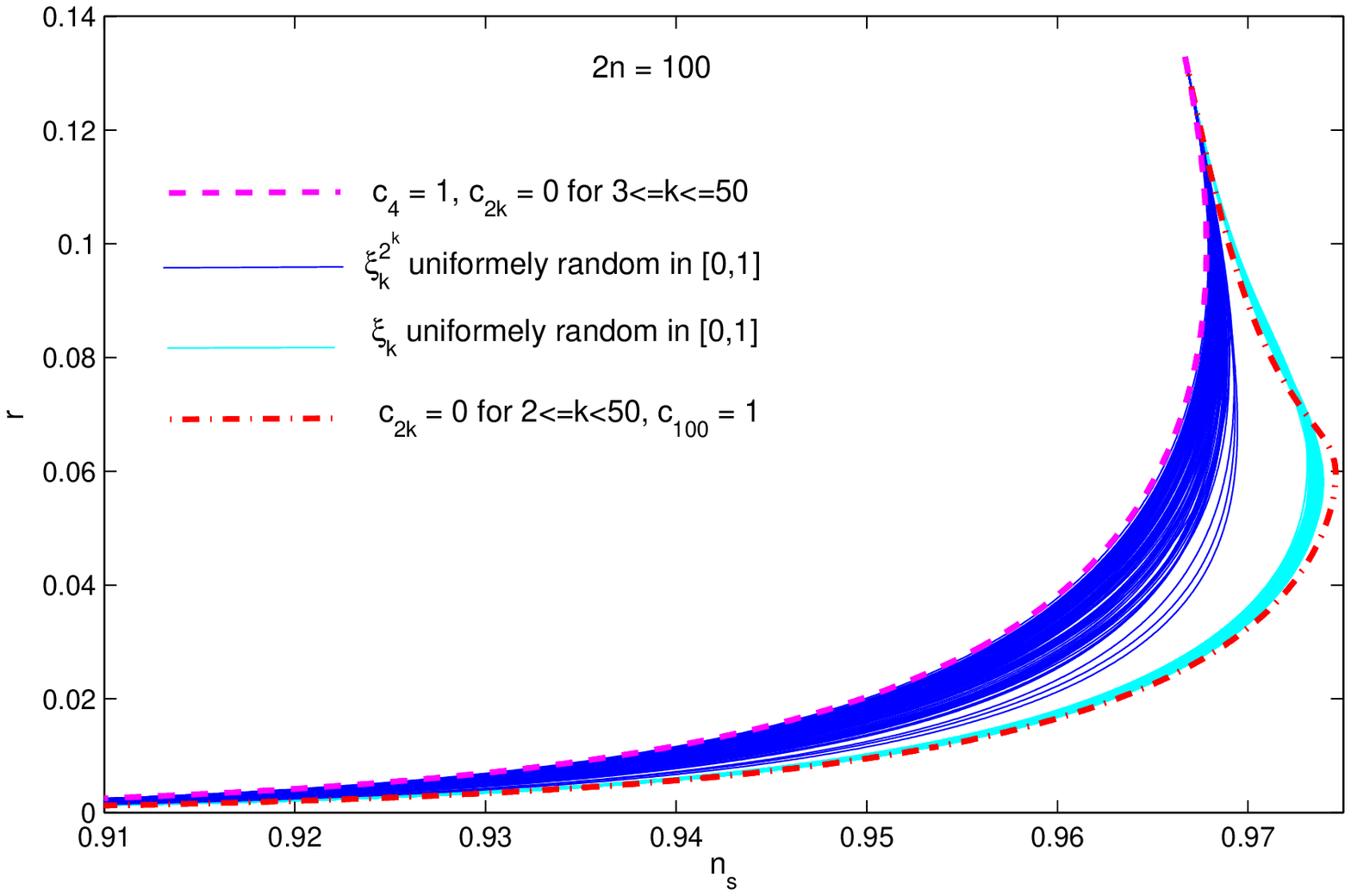}
\includegraphics[height=10cm]{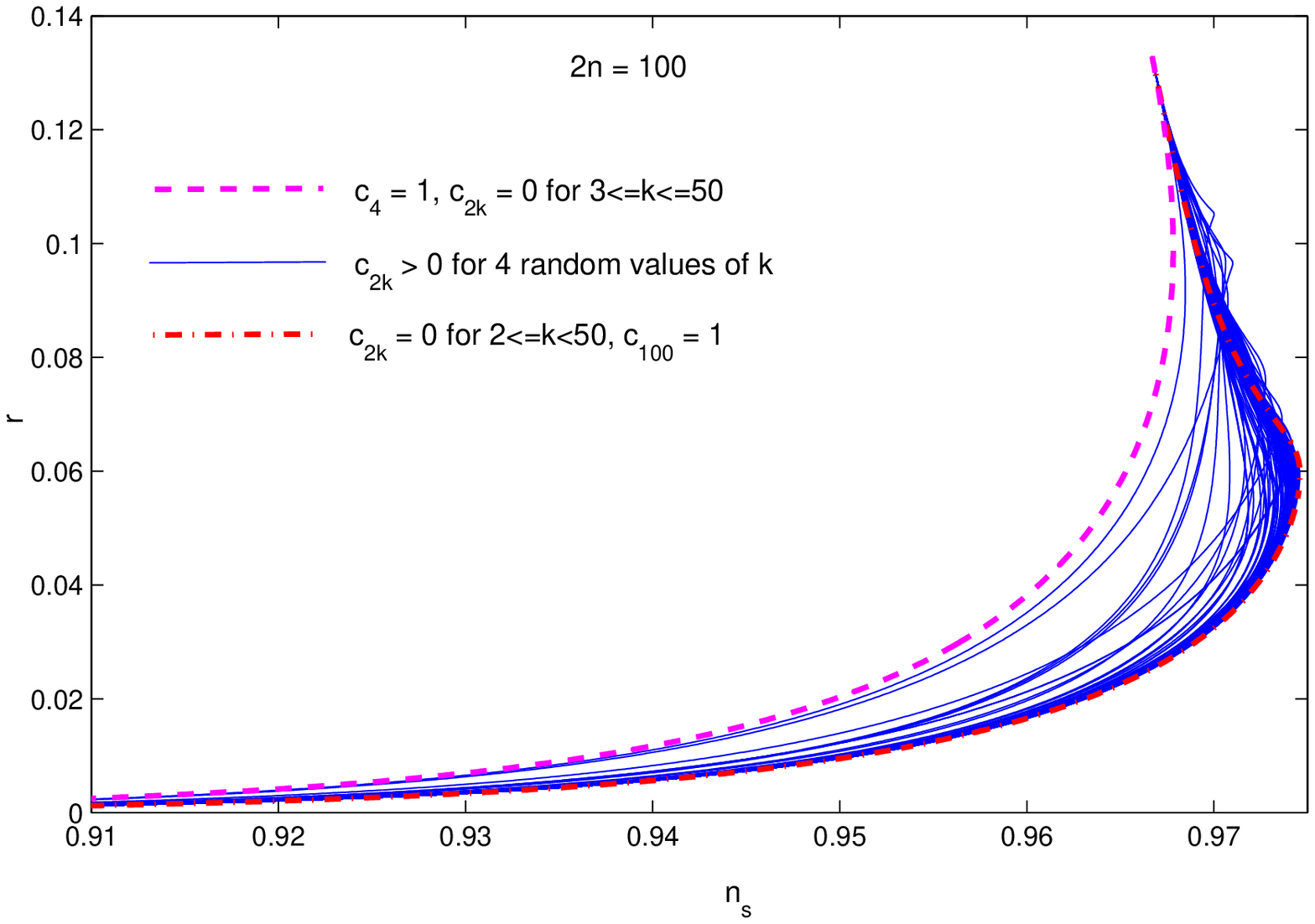}
\caption{$ r $ vs. $ n_s $ for the 100th. order polynomial potential
eq.(\ref{eq:vg2n}) for $ n = 50 $. The coefficients $ c_{2k} $ were chosen 
or extracted at random as indicated in the two panels. The two border
curves of the banana region $ \cal B $ are clearly indicated. The upper
border is the fourth order potential eq.(\ref{binon}) and the lower border 
is the quadratic plus the $2n$th order potential eq.(\ref{wn}).
The quadratic plus the $2n$th order potential always provides the lowest 
value for $ r $ at any fixed $ n_s $ in its lower branch. In the upper 
panel, all the coefficients $ c_{2k} $ were extracted 
independently from a flat distribution ranging from 0 to 100; in this case 
the curves {\bf accumulate} near the quadratic plus 
quartic potential eq.(\ref{binon}). The upper panel is the {\bf generic} case.
In the lower panel, we picked at random four possibly non-zero 
$ c_{2k} $ and fixed to zero the remaining 44 ones; in this case 
the curves {\bf accumulate} near the quadratic plus $2n$th order potential 
eq.(\ref{wn}) with $ 2 \, n = 100 $.}
\label{fig:ban100}
\end{figure}

For higher values of $ n $ we extracted the numbers $ \xi_k $ at random within the unit
interval. In particular, for the highest case considered, $ n=50 $, we used three
distributions: in the first, the $ \xi_k $ were all extracted independently and
uniformly over the unit interval; in the second we set 
$ \log\xi_k = 2^{-k} \; \log{\tilde\xi_k} $
and extracted the $ {\tilde\xi_k} $ independently and uniformly; in the third we
picked at random four $ \xi_k $ freely varying and fixed to 1 the remaining 45
ones (that is we picked at random four possibly non--zero $ c_{2k} $, setting the
rest to zero); the values of the four free $ \xi_{k} $ were chosen at random in
the same set of values $ (0.001,\;0.5,\;0.999) $ of the $ n=5 $ case.  The results
of these simulations are shown in fig.~\ref{fig:ban100}.

As evident from fig.~\ref{fig:splitban10}, where the $ r=r(n_s) $ curves are split
in upper/lower branches with growing/decreasing $ r=r(n_s) $ and especially from
fig.~\ref{fig:ban100}, the case of the quadratic plus $ 2n$th order polynomial
provides a bound to the banana region $ {\cal B} $ from below. That is, 
for any fixed value of $ n _s $, the quadratic plus $ 2n$th order
polynomial provides the {\bf lowest} value for $ r $.

One sees from fig. \ref{fig:ban100} that some blue curves $ r=r(n_s) $ go beyond
the slashed red curve $ r=r(n_s) $ for the quadratic plus $ u^{100} $ potential
on the right upper border of the banana region $ {\cal B} $. Namely, the 
right upper border of the $ {\cal B} $ region is not given by the 
quadratic plus $ u^{100} $ potential while this potential provides the
lower border of the $ {\cal B} $ region.

\medskip

We performed many other tests with intermediate values of $ n $ and several 
other distributions, including other $k-$dependent distributions, with 
characteristic values
for $ c_{2k} $ growing linearly with $ k $ or decreasing in a power--like or
exponential way. In all cases, the results were consistent with those
given above.
 
\begin{figure}[h]
\includegraphics[width=12cm]{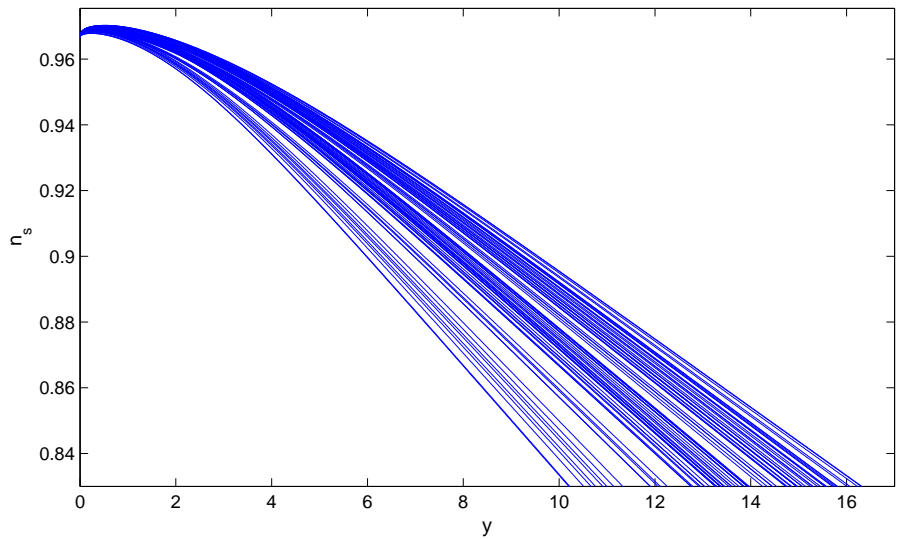}
\caption{$ n_s $ vs. the coupling $ y $ within the same setup as in 
fig.~\ref{fig:ban10}.}
\label{fig:nsvy10}
\end{figure}

It is also important to observe that the class of potentials considered, that is
arbitrary even polynomials with positive or zero couplings, is a class of weakly
coupled models. This is evident from fig.~\ref{fig:nsvy10}, were $ n_s $ is
plotted vs. the coupling $ y $, which remains of order one when $ n_s $ decreases
well below the current experimental limits. This weak coupling is the reason why
the addition of higher even monomials to these potentials causes only minor
quantitative changes to the shape of the $ r=r(n_s) $ curves.

The inflaton potential $ V(\varphi) $ eq.(\ref{eq:vg2n}) in the original
inflaton field $ \varphi $ takes therefore the form
$$
 V(\varphi) = \frac{4 \, N \; M^4}{y(u)} \; \left\{ 1-  \frac{y(u)}{8 \, N} \;
\frac{\varphi^2}{M_{Pl}^2}+\sum_{k=2}^{n}\frac{c_{2k}}{k} \;
\left(\frac{y(u)}{8 \, N}\right)^k \; \frac{\varphi^{2 \, k}}{M_{Pl}^{2\,k}} \right\}\; ,
$$
Therefore, since the coupling $ y(u) $ is $ {\cal O}(1) $ we have
the $2 \, k$-th term in the potential suppressed by the $2 \, k$-th 
power of $ M_{Pl} $ as well as by the factor $ N^k \sim 60^k $. 

In particular, the quartic term 
$$
\frac{y(u) \; c_4}{32 \, N} \left(\frac{M}{M_{Pl}}\right)^4 \; \varphi^4
$$
possesses a very small quartic coupling since $ M \ll M_{Pl} $. 
Notice that these suppression factors are natural in the GL approach
and come from the ratio of the two relevant energy 
scales here: the Planck mass and the inflation scale $ M $.
When the GL approach is not used these suppression factors do not follow in general.

The validity of the GL approach relies on the wide separation between the scale of 
inflation and the higher energy scale $ M_{Pl} $ (corresponding to the 
underlying unknown microscopic theory as discussed in  ref. \cite{1sN}.) 
It is not necessary to require $ \varphi \ll M_{Pl} $ in the GL approach
but to impose \cite{1sN}
$$
V(\varphi) \ll M_{Pl}^4 \quad {\rm and ~hence} \quad  
v_1\left(\phi \; \sqrt{g}\right) \ll 10^{12} \; g \; .
$$
This last condition gives an upper bound for the inflaton field $ \varphi $
depending on the large argument behavior of $ v_1(u) $. We get for example:
$$ 
\varphi \ll 10^6 \;  M_{Pl} \quad {\rm  for} \quad   v_1(u) \buildrel{u \to \infty}\over\sim u^2 
\quad ,  \quad  \varphi \ll 2600  \;  M_{Pl} \quad {\rm  for} \quad   
v_1(u) \buildrel{u \to \infty}\over\sim u^4 \; .
$$
The validity of the effective GL theory relies on that separation of scales 
and the GL approach allows to determine the scale of inflation as $ 0.543 \times 10^{16} $ GeV 
(at the GUT scale) and well below the Planck scale  $ M_{Pl} $
using the amplitude of the scalar fluctuations from the CMB data \cite{mcmc,bibl}.

Inflaton potentials containing terms of arbitrary high order in
the inflaton are considered in ref. \cite{lily}, sec. 25.3.2
without using the GL approach and within the small field hypothesis 
$ \varphi \ll M_{Pl} $. Smallness conditions on the expansion coefficients 
are required in ref. \cite{lily}. This is actually not needed in the GL approach,
whose validity relies only on the wide separation of scales between $ M $ and $ M_{Pl} $,
at least in the case of even polynomials with positive coefficients.

\begin{figure}[h]
\begin{turn}{-90}
\psfrag{"nsrqq6.dat"}{$n=3$}
\psfrag{"nsrn2.dat"}{$n=2$}
\psfrag{"nsrn.dat"}{$n=5$}
\psfrag{"nsrn10.dat"}{$n=10$}
\psfrag{"nsrn20.dat"}{$n=20$}
\psfrag{"nsrn100.dat"}{$n=100$}
\psfrag{"nsrinf2.dat"}{$n=\infty$}
\psfrag{"nsrinf.dat"}{$n=\infty, \; u = 1$}
\includegraphics[height=15cm,width=10cm]{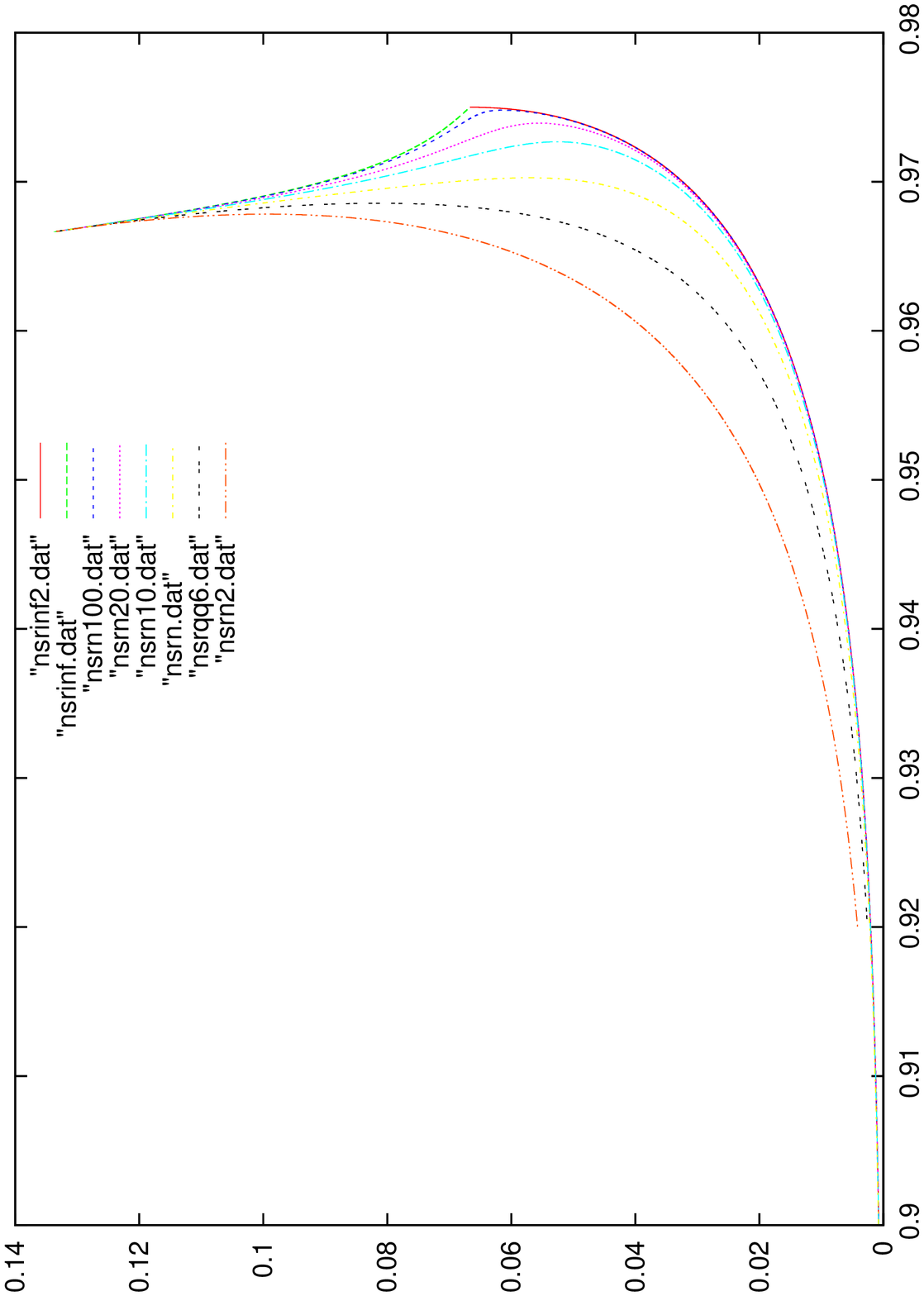}
\end{turn}
\caption{$ r $ vs. $ n_s $ for the quadratic plus $ u^{2 \, n} $ potential 
eq.(\ref{bini}) setting $ N = 60 $. The curves for the exponents 
$ n = 2, \; 3, \; 5, \; 10,  \; 20 $ and $ 100 $ are displayed as well as 
the limiting curves obtained in the $ n = \infty $ 
limits eqs.(\ref{nsrinf}) and (\ref{rnsinf}). Eq.(\ref{nsrinf}) describes the
lower bordering curve while eq.(\ref{rnsinf}) describes the upper-right
 bordering curve. We see that for growing 
$ n $ the curves $ r $ vs. $ n_s $ tend towards the {\bf limiting} curves.
The uppermost point where all curves coalesce 
corresponds to the monomial quadratic potential 
$ n_s = 0.9666\ldots, \; r =  0.13333 \ldots $ [see eq.(\ref{cobino})].}
\label{nsrseisn}
\end{figure}

\section{The quadratic plus the $ 2 \, n $th order double-well inflaton 
potential}\label{pot2n}

In order to find the observationally interesting right and down border of 
the banana we consider the quadratic plus the $ 2 \, n $th order
potential for new inflation \cite{ho},
\be\label{wn}
v_1(u) = \frac12 \; \left(1-u^2\right) + \frac1{2 \; n} \; 
\left(u^{2 \, n}-1 \right) \; .
\ee
As in the general case eq.(\ref{v1gen}), we choose the absolute minimum at 
$ u = 1 $. The customary relation eq.(\ref{eq:gN}) takes here the form 
\cite{ho},
\be\label{dos}
y(u) = \frac4{n} \; 
\int_u^1 \frac{dx}{x}\frac{n \; (1-x^2) + x^{2 \, n} -1}{1- x^{2 \, n-2}}
\quad {\rm where} \quad  0< u <1 \; .
\ee
This integral can be expressed as a sum of $ n $ terms including 
logarithms and arctangents \cite{PBM}. 

In the weak coupling limit $ y \to 0, \; n_s $ and $ r $ take the 
values of the quadratic monomial potential 
eqs.(\ref{eq:rweak})-(\ref{eq:nsweak}) \cite{ho,bibl}:
\be\label{gva0}
n_s - 1\buildrel{y \to 0}\over= - \frac2{N}=-0.0333\ldots \quad , \quad
r \buildrel{y \to 0}\over= \frac8{N}=0.1333\ldots\quad ,
\ee
while in the strong coupling limit $ y \to \infty $ at 
fixed $ n , \; n_s $ and $ r $ take the values 
$$
n_s \simeq 1 + \frac2{N} \; \log u  \longrightarrow -\infty \quad ,\quad
r \simeq -\frac{16}{N} \; \frac{n}{n-1} \; u^2 \; \log u \longrightarrow 0^+
  \; ,
$$
in accordance with the general formula eq.(\ref{eq:strong}). 
In fig. \ref{nsrseisn} we plot $ r $ vs. $ n_s $ for the potential
eq.(\ref{wn}) and the exponents 
$ n = 5, \; 10,  \; 20, \; 100, \; 500 $ and $ 5000 $. We see that for 
$ n \to \infty , \; r $ vs. $ n_s $ tends towards a {\bf limiting} curve. 
For $ y \to 0 $ we reach the upper end of the curve 
[the monomial quadratic potential eq.(\ref{gva0})] while for large 
$ y $ the left and lower end of the curve is reached. However,
the current CMB--LSS data rule out this strong coupling part of the curve
for $ n_s < 0.95 $.

\subsection{The $ n \to \infty $ limit at fixed $ u $.}\label{ninfXf}

Let us first compute $ y(u) $ eq.(\ref{dos}) for $ n \to \infty $ 
at {\bf fixed} $ u $. Since $ 0 < x <1 $ in the integrand of eq.(\ref{dos}), 
$$
{\displaystyle \lim_{n \rightarrow \infty}} x^{2 \, n} = 0 \; .
$$
and eq.(\ref{dos}) reduces to
$$
y(u) \buildrel{n \to \infty}\over=  \frac4{n} \; 
\int_u^1 \frac{dx}{x} \; \left[n \; (1-x^2) -1\right]=
2 \left[u^2-1-\ln u^2 + {\cal O}\left(\frac1{n}\right)\right]  \; .
$$
Hence, eq.(\ref{dos}) becomes
\be\label{ginf2}
y(u) \buildrel{n \to \infty}\over= 2 \; \left( -\ln u^2  -1 + u^2\right)
\quad {\rm where} \quad 0< u <1 \quad {\rm and} \quad 0< y < +\infty \; .
\ee
which is just twice the result found in the quartic double--well potential,
eq.~(\ref{eq:yexpl4}). Notice that $ v_1(u) $ eq.(\ref{wn}) in the $ n
\to \infty $ limit becomes 
\be\label{limv}
{\displaystyle \lim_{n\to \infty}} v_1(u) = \left\{
\begin{array}{l}\frac12 \; (1-u^2) \quad {\rm for} \; u<1 \\
  +\infty \quad {\rm for} \;  u>1 \; .
       \end{array} \right.\; .
\ee
From eqs.(\ref{eq:nsr}), (\ref{wn}) and (\ref{ginf2})
we find for $ r $ and $ n_s $ in the $ n \to \infty $ limit
\bea\label{nsrinf}
&& n_s - 1 \buildrel{n \to \infty}\over= -\frac1{N} \; 
\frac{2 \; u^2+1}{(1-u^2)^2} \; \left( -\ln u^2  -1 + u^2\right) \; , \cr \cr 
&& r \buildrel{n \to \infty}\over= \frac8{N} \; \frac{u^2}{(1-u^2)^2} \; 
\left( -\ln u^2  -1 + u^2\right) \; .
\eea
Now, in the limiting cases $ u \to 0 $ and $ u \to 1 $
({\bf at} $ n = \infty $), that is, the strong coupling limit $ y \to \infty $ and the 
the weak coupling limit $ y \to 0 $, 
respectively, we obtain from eqs.(\ref{nsrinf}) 
\bea\label{ptoX}
&&
 {\displaystyle \lim_{u \rightarrow 1}} \; n_s(n = \infty) -1 =  -\frac3{2 \, N} 
= - \frac1{40} = -0.025 \quad , \quad
{\displaystyle \lim_{u \rightarrow 1}} \; r(n = \infty) = \frac4{N} = 
\frac1{15} = 0.0666\ldots \; , \\ \cr
&& {\displaystyle \lim_{u \rightarrow 0}} \; n_s(n = \infty) = -\infty\quad , \quad 
{\displaystyle \lim_{u \rightarrow 0}} \; r(n = \infty) =0 \; . \nonumber
\eea
However, as explained in sec. \ref{sec2}, the slow-roll 
expansion is no more valid when $ | n_s - 1 | \gtrsim 1 $. Moreover, 
the WMAP+LSS results exclude $ n_s \lesssim 0.9 $ \cite{WMAP5}. 
Therefore, the limit $ u \to 0 $ is ruled out.

\medskip

Eqs.(\ref{nsrinf}) describe the rightmost (limiting) curve in fig. 
\ref{nsrseisn} in its {\bf lower} part, namely $ 0 < r < 4/N = 
0.0666\ldots $. The upper part is obtained in the double limit 
$ n \to \infty $ {\bf and} $ u \to 1 $ (or, equivalently
$ n \to \infty $ {\bf and} $ y \to 0 $), as we show in the next section.

\subsection{The double limit $ n \to \infty 
$ {\bf and} $ u \to 1 $.}\label{nXinf}

As we can see from fig. \ref{nsrseisn}, when $ y $ varies from zero to 
infinity at fixed $ n $, the potential eq.(\ref{wn}) covers the region
$$
0 < r < \frac8{N} \; ,
$$
the point $ r = 8/N $ corresponding to the small coupling 
limit  $ y = 0 $. 

Notice however that the $ n \to \infty $ limit eqs.(\ref{ginf2}) and 
(\ref{nsrinf}) {\bf only} describe the region 
$ 0 < r < 4/N $. In order to also describe the small coupling region 
$ 8/N > r > 4/N $ for $ n \to \infty $, we have to take in eq.(\ref{dos})
the {\bf double} limit $ u \to 1 $ {\bf and} $ n \to \infty $. 
This can be achieved by changing the integration variable in 
eq.(\ref{dos}) as $ x = t^{\frac1{2n}} $,
$$
y(u) = \frac2{n^2}\int_\tau^1\frac{dt}{t}\frac{n \; 
(1-t^{\frac1{n}}) + t -1}{1- t^{1-\frac1{n}}}
\quad {\rm where}\quad \tau \equiv u^{2 \, n} \; , \quad 0 < \tau < 1 \quad.
$$
Letting $ n \to \infty $ at {\bf fixed} $ \tau $ yields,
\be\label{tres}
\frac{n^2}2 \; y(u) \buildrel{n \to \infty, \;  u \to 1}\over= 
\int_\tau^1\frac{dt}{t}\frac{t-1-\ln t}{1-t}= \ln \tau + \frac12 \; 
\ln^2 \tau +{\rm Li}_2(1-\tau) \; ,
\ee
where
$$
{\rm Li}_2(s) = -\int_0^s \frac{dt}{t} \; \ln(1-t) \; ,
$$
is the dilogarithmic function \cite{PBM}.

Then, in this {\bf double limit} $ n \to \infty , \;  u \to 1 $  
eq.(\ref{dos}) becomes 
\be\label{ginf}
\gamma^2(\tau) \equiv \frac{n^2}2 \; y(u) \buildrel{n \to \infty, \;  
u \to 1}\over= \ln \tau + \frac12 \; \ln^2 \tau  +{\rm Li}_2(1-\tau) \; .
\ee
That is,  $ \tau $ and $ \gamma^2 $ {\bf are fixed}
in this $ n \to \infty , \;  u \to 1 $ limit.  
Notice that $ 0 < \tau < 1, \; 0 < \gamma < \infty $ while 
$ y \to 0 $
\be\label{limni}
y(u) \buildrel{n \to \infty, \; u \to 1}\over= \frac{2 \; \gamma^2(\tau)}{n^2} \to 0 
\quad {\rm and} \quad 
u = \tau^{\frac1{2 \, n}}\buildrel{n \to \infty}\over= 1 +  
{\cal O}\left(\frac1{n}\right) \; .
\ee
From eq.(\ref{eq:nsr}) the spectral index $ n_s $, and the ratio of tensor 
to scalar fluctuations $ r $ for fixed $ \gamma $ and $ \tau $
take here ($ n = \infty, \; y = 0 $ and  $ u = 1 $) the following form,
\bea\label{rnsinf}
&& n_s - 1 \buildrel{n \to \infty, \;  u \to 1}\over= 
-\frac{3 \, \gamma^2(\tau)}{N} \frac{(1-\tau)^2}{(\tau-1-\ln \tau)^2} + 
\frac{2 \, \gamma^2}{N} \frac{\tau}{\tau-1-\ln \tau} \; ,  \cr \cr
&& r \buildrel{n \to \infty, \;  u \to 1}\over= \frac{8 \, \gamma^2(\tau)}{N} 
\frac{(1-\tau)^2}{(1-\tau+\ln \tau)^2} \; .
\eea
We obtain from eqs.(\ref{rnsinf}) in the limiting cases $ \tau \to 0 $ and 
$ \tau \to 1 $,
\bea\label{ptoT}
&& 
{\displaystyle \lim_{\tau \rightarrow 0}} n_s -1 = -\frac3{2 \, N} 
= - \frac1{40}= -0.025 \quad , \quad  {\displaystyle \lim_{\tau \rightarrow 0}} r 
=\frac4{N} = \frac1{15} = 0.0666\ldots \; , \cr \cr
&&  {\displaystyle \lim_{\tau \rightarrow 1}} 
n_s -1 = - \frac2{N} =  - \frac1{30} = -0.0333\ldots \quad , \quad 
{\displaystyle \lim_{\tau \rightarrow 1}} r = \frac8{N} = \frac2{15} 
= 0.1333\ldots 
\eea
Notice that $ n_s $ and $ r $ for $ n \to \infty $ and {\bf then} 
$ u \to 1 $ eq.(\ref{ptoX}) coincides with $ r $ and  $ n_s $ in the 
double limit $ n \to \infty , \; u \to 1 $ for  $ \tau = u^{2 \, n} \to 0 $ 
eq.(\ref{ptoT}). Namely, eqs.(\ref{nsrinf}) and (\ref{rnsinf}) match to 
each other as eqs.(\ref{ptoX}) and (\ref{ptoT}).

\medskip

Eqs.(\ref{rnsinf}) describe the rightmost (limiting) curve in fig. 
\ref{nsrseisn} in its {\bf upper} part, namely $ 8/N = 0.1333\ldots 
> r > 4/N = 0.0666\ldots $. 
The lower part, $ 0 < r < 4/N = 0.0666\ldots $, is described by 
eqs.(\ref{nsrinf}).  $ r $ and $ n_s $ 
given by eqs.(\ref{nsrinf}) and (\ref{rnsinf}) continuously match 
at $ n_s = 0.975 , \; r = 0.0666\ldots $. However, the derivative 
$ dr/dn_s $ is discontinuous at this point. 

There is here a quadratic relation between $ n_s $ and $ r $ for 
$ r \to 4^-/N $ valid in the $ n = \infty $ limit:
\be\label{relq}
\left(r - \frac4{N}\right)^2 = - \frac{64}{3 \, N}\left(n_s - 1 + 
\frac3{2 \, N} \right) \left[1 + {\cal O}\left(\sqrt{n_s - 1 + 
\frac3{2 \, N}}\right) \right] \; .
\ee
From eqs.(\ref{relq}) and (\ref{rnsinf}) we get respectively
$$
{\displaystyle \lim_{r \to 4^-/N}} \; \frac{dr}{dn_s} = + \infty
\quad , \quad {\displaystyle \lim_{r \to 4^+/N} } \; \frac{dr}{dn_s} = - \frac83 \; ,
$$
as we can see in  fig. \ref{nsrseisn}.

\section{The quadratic plus the exponential potential.} \label{potex}

Since the exponential function contains all powers of the variable, it is 
worthwhile to consider it. As before, we restrict ourselves to potentials 
even in $ u $:
\be\label{expo}
 v(\phi) = \frac{c_0}{\hat g} - \frac12  \; \phi^2 + 
\frac1{2 \, {\hat g} \; c}\left(e^{ {\hat g} \; \phi^2} - 1 - 
{\hat g} \; \phi^2\right)  \; ,
\ee
where $ {\hat g} > 0 $ and $ c > 0 $ are free parameters, while as usual
$ c_0 $ ensures that $ v(\phi) $ vanishes at its absolute minimum 
$ \phi = \phi_{min} = 1/\sqrt{g} $. We find
$$
\phi_{min} = \frac1{\sqrt{g}} =\sqrt{\frac1{{\hat g}} \log(1 + c )} \quad , \quad 
  b \equiv \frac12 \; \log(1 + c) > 0 \quad , \quad g = \frac{\hat g}{2 \, b} 
$$
and 
$$
c_0 = \frac12 \left[ \left(1 + \frac1{c}\right)\log(1 + 
c)-1 \right]
$$
In terms of the variable $ u = \phi/\phi_{min} $ the potential $ v_1(u) $
defined in general by eq.(\ref{eq:vg}) takes here the form,
\be\label{v1ex}
v_1(u) = \frac{e^{-2 \, b \; (1-u^2)} - 1+ 
2 \, b \; (1-u^2)}{4 \, b \left(1-e^{-2 \, b}\right)} \; .
\ee
Expanding the potential eq.(\ref{v1ex}) in powers of $ u $ yields
$$
v_1(u) \buildrel{u \to 0}\over= \frac{1 + e^{2 \, b} \; (2 \, b-1)}{4 \, b \; 
( e^{2 \, b}-1)} - \frac12 \; u^2 + \frac{b}{2 \; (e^{2 \, b}-1)}  
\; u^4 + {\cal O}(u^6) \; .
$$
It is interesting to expand the potential in powers of $ b $ in order to make 
contact with the polynomial potentials of sec. \ref{sec:dbw}-\ref{sextico}.
We get from eq.(\ref{v1ex})
$$
v_1(u) \buildrel{b \to 0}\over=\frac1{12} \,(1-u^2)^2\,(3+b+2\,b\,u^2) 
  + {\cal O}(b^2)   
$$
which is exactly the fourth order double--well potential eq.(\ref{binon})
to zeroth order in $ b $ and the sixth--order double--well potential 
eq.(\ref{v1s}) to first order in $ b $.

\medskip

The field $ u $ at horizon exit follows from the customary eq.(\ref{eq:gN})
which takes here the form:
\be\label{unex}
y(u) = \frac2{b} \int_1^{u} \frac{dx}{x} \; 
\frac{e^{-2 \, b \; (1-x^2)} + 2 \, b \; (1-x^2) - 1}{e^{-2 \, b \; (1-x^2)} - 1} \; ,
\ee
Changing the integration variable to $ w \equiv 1 - e^{-2 \, b \; (1-x^2)} $, 
eq.(\ref{unex}) becomes
\be\label{4bg}
y(u) = 2 \; \left( -\ln u^2  -1 + u^2\right) + \frac2{b} \; \ln u
-\frac1{b}\int_0^{1 - e^{-2 \, b \; (1-u^2)}} \frac{dw}{w} \; 
\frac{\log(1-w)}{2 \, b + \log(1-w)} \; .
\ee
The spectral index $ n_s $,  and the ratio $ r $ are expressed from
eq.(\ref{eq:nsr}) as, 
\bea\label{bnsr}
&& n_s - 1 = -\frac38 \; r + \frac{b \; y(u)}{N} 
\frac{(4 \, b \; u^2 + 1)
\;  e^{-2 \, b \; (1-u^2)}-1}{e^{-2 \, b \; (1-u^2)} + 2 \, b \; (1-u^2)- 1} \; , \cr \cr
&& r = \frac{16 \, b^2}{N} \; u^2 \; y(u) \; 
\left[\frac{ e^{-2 \, b \; (1-u^2)} - 1}{ e^{-2 \, b \; (1-u^2)} + 2 \, b \; (1-u^2)
- 1} \right]^2 \; .
\eea
We study below eqs.(\ref{4bg})-(\ref{bnsr}) in the $ b \to \infty $ limit 
in the two regimes: $ b \to \infty $ with $ u $ {\bf fixed}
and $ b \to \infty $ {\bf with} $ u \to u_{min} = 1 $. 
These are the limits investigated
in secs. \ref{ninfXf} and \ref{nXinf} for the quadratic 
plus $ u^{2 \, n} $ potential, respectively.

\begin{figure}[h]
\begin{turn}{-90}
\psfrag{"nsr4.dat"}{$b=0$}
\psfrag{"bensr1.dat"}{$b = 0.8 $}
\psfrag{"bensr2.dat"}{$b = 4 $}
\psfrag{"bensr3.dat"}{$b = 8 $}
\psfrag{"bensr4.dat"}{$b = 40 $}
\psfrag{"bensr6.dat"}{$ b = 160 $}
\psfrag{"bensr9.dat"}{$ b = 1600 $}
\psfrag{"bensrX.dat"}{$ b = 1.6 \times 10^5$}
\psfrag{"bensrX6.dat"}{$ b = 8 \times 10^8 $}
\psfrag{"nsrinf.dat"}{$ b = \infty , \; u = 1 $}
\psfrag{"nsrinf2.dat"}{$ b = \infty $}
\includegraphics[height=15cm,width=10cm]{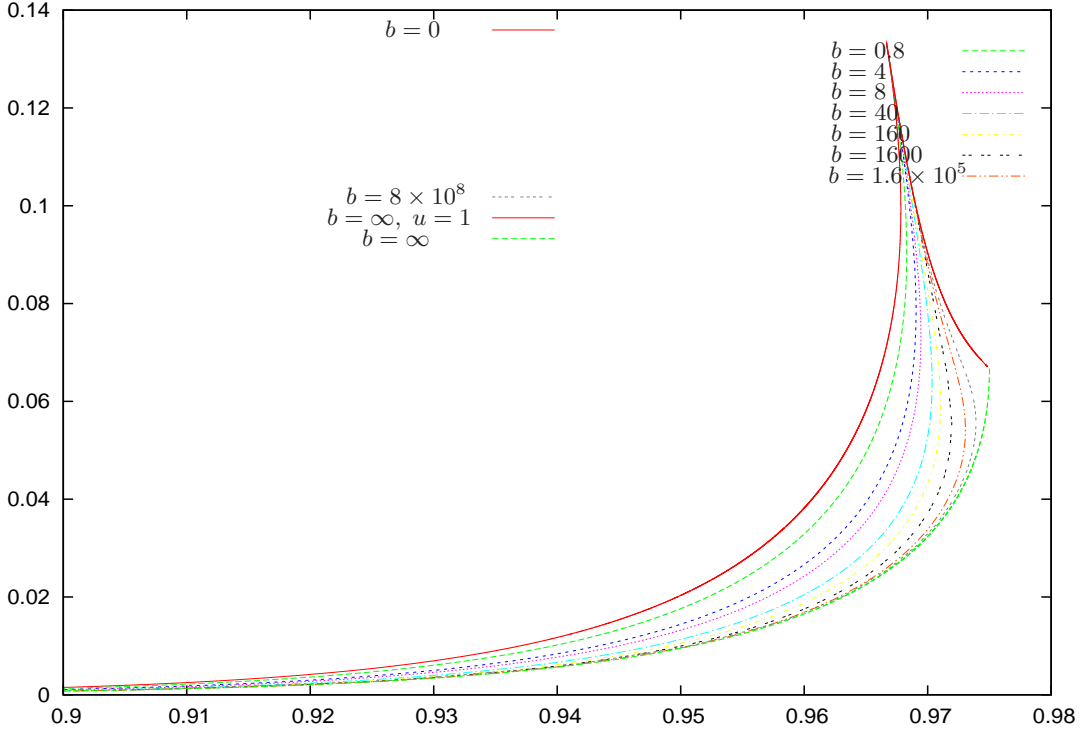}
\end{turn}
\caption{$ r $ vs. $ n_s $ for the quadratic plus exponential potential
eq.(\ref{expo}) with the coefficient $  0 \leq b \leq \infty $ and setting 
$ N = 60 $. We see that for growing $ b \gg 1 $, $ r $ vs. $ n_s $ tends 
towards a {\bf limiting curve} to the right and down of the banana shaped
region $ \cal B $. This curve is the lower border of the region $ \cal B $.
The upper border is determined by the fourth order potential 
eq.(\ref{binon}). The uppermost point where all curves coalesce corresponds 
to the monomial quadratic potential $ n_s = 0.9666\ldots, \; 
r =  0.13333 \ldots $ [see eq.(\ref{cobino})].}
\label{expnsr}
\end{figure}

In fig. \ref{expnsr} we plot $ r $ vs. $ n_s $ for the quadratic plus
exponential potential eq.(\ref{expo}) and the values of the coefficient 
$ c = 0.1, \; 0.5,  \; 1, \; 5, $ and $ 10 $. We see that for growing 
$ c ,\;  r $ vs. $ n_s $ tends towards a {\bf limiting} curve. This curve
is the {\bf lower} border of the banana shaped region $ \cal B $. The upper
border is determined by the fourth order potential eq.(\ref{binon}).

\subsection{The limit $ b \to \infty $ at fixed $ u $.}\label{binfX}

For large $ b $ we have in eqs.(\ref{4bg})-(\ref{bnsr}),
$$
e^{-2 \, b \; (1-u^2)} \ll 1 \quad {\rm since} \quad  b \gg 1 \quad {\rm and} 
\quad u < 1 \; ,
$$
and we find
\bea
&& v_1(u) \buildrel{b \to \infty}\over= \frac12 \; (1-u^2)+ 
{\cal O}\left(\frac1{b}\right)\quad {\rm for} \; u<1 \; ,  \cr \cr
&& {\displaystyle \lim_{n\to \infty}} v_1(u) = +\infty 
\quad {\rm for} \; u>1 \; ,  \cr \cr
&&  y(u) \buildrel{b \to \infty}\over= 2 \; 
\left( -\ln u^2  -1 + u^2\right)
+ {\cal O}\left(\frac1{b}\right) \; ,  \cr \cr
&& n_s - 1 \buildrel{b \to \infty}\over= -\frac1{N} \; 
\frac{2 \; u^2+1}{(1-u^2)^2} \; \left( -\ln u^2  -1 + u^2\right) 
+ {\cal O}\left(\frac1{b}\right) \; , \cr \cr 
&& r \buildrel{b \to \infty}\over= \frac8{N} \; \frac{u^2}{(1-u^2)^2} \; 
\left( -\ln u^2  -1 + u^2\right) + {\cal O}\left(\frac1{b}\right) \; .
\eea
These equations for  $ r $ vs. $ n_s $ exactly {\bf coincide} with 
eqs.(\ref{limv})-(\ref{nsrinf}) for the quadratic plus $2n$th order 
potential. We have therefore proved that the 
quadratic plus the $ u^{2 \, n} $ potential and the quadratic plus 
exponential potential have {\bf identical} limits letting 
$ n \to \infty $ in the former and $ b \to \infty $ in the latter, 
keeping always $ u $ fixed.

\subsection{The double limit $ b \to \infty $ {\bf and} $ u \to 1 $.}

It is useful to introduce here the variable
$$
\tau \equiv  e^{-2 \, b \; (1-u^2)}
\quad  {\rm hence} \quad u^2 = 1 +
\frac{\log \tau}{2 \, b} \to 1^-
\quad {\rm for} \quad b \to \infty \quad {\rm at ~fixed} \; \tau \quad , \quad
0 < \tau < 1 \; .
$$
We then find from eq.(\ref{4bg}) for $ b \to \infty $ and fixed $ \tau $,
\bea
&& 2 \; b^2 \; y(u) = 2 \; \left( -\ln u^2  -1 + u^2\right) + \frac2{b} \; \ln u
-\frac1{2 \, b^2}\int_0^{1 - \tau} \frac{dw}{w} \; \log(1-w)  +  
{\cal O}\left(\frac1{b}\right) = \cr \cr
&& = \ln \tau + \frac12 \; \ln^2 \tau +{\rm Li}_2(1-\tau) + 
{\cal O}\left(\frac1{b}\right) \; ,
\eea
We find in this limit from eq.(\ref{bnsr}) for $ r $ vs. $ n_s $,
\bea\label{rnsU}
&& r \buildrel{b \to \infty, \;  u \to 1}\over= \frac{8 \, \gamma^2(\tau)}{N} 
\frac{(1-\tau)^2}{(1-\tau+\ln \tau)^2} \; , \cr \cr
&& n_s - 1 \buildrel{b \to \infty, \;  u \to 1}\over= 
-\frac{3 \, \gamma^2(\tau)}{N} \frac{(1-\tau)^2}{(\tau-1-\ln \tau)^2} + 
\frac{2 \, \gamma^2(\tau)}{N} \frac{\tau}{\tau-1-\ln \tau} \; , \cr \cr
&& \gamma^2(\tau) \equiv 2 \; c^2 \; y(u)
\buildrel{b \to \infty, \;  u \to 1}\over=
 \ln \tau + \frac12 \; \ln^2 \tau + {\rm Li}_2(1-\tau) \; ,
\eea
where we keep {\bf fixed} $ \gamma^2 $. 
Eqs.(\ref{rnsU}) {\bf coincide} with eqs.(\ref{ginf})-(\ref{rnsinf}) for 
the quadratic plus $ u^{2 \, n} $ potential.

These results plus those in sec. \ref{binfX} {\bf prove} that the 
quadratic plus $ u^{2 \, n} $ potential and the quadratic plus 
exponential potential eq.(\ref{expo}) have {\bf identical} limits letting 
$ n \to \infty $ in the former and $ b \to \infty $ in the latter.

\section{Dynamically generated inflaton potential from a fermion condensate
 in the inflationary stage.} \label{potferm}

The inflaton may be a coarse-grained average of
fundamental scalar fields, or a composite (bound state) or condensate of 
fields with spin, just as in superconductivity. Bosonic fields do
not need to be fundamental fields, for example they may emerge as
condensates of fermion-antifermion pairs $ < {\bar \Psi} \Psi> $
in a grand unified theory (GUT) in the cosmological background 
\cite{bibl}.

We investigate in this section an inflaton potential dynamically generated
as the effective potential of fermions in the inflationary universe.
We consider the inflaton field coupled to Dirac fermions $ \Psi $ through 
the interaction
Lagrangian
\be\label{Lf}
{\cal L} = \overline{\Psi}\left[i\,\gamma^\mu \;  \mathcal{D}_\mu  -m_f -
g_Y \; \varphi \right]\Psi \; .
\ee
Here $ g_Y $ stands for a generic Yukawa coupling between the fermions and
the inflaton $ \varphi $. The fermion mass $ m_f $ can be absorbed in a
constant shift of the inflaton field. The Dirac matrices $ \gamma^\mu $ are the curved 
space-time $ \gamma $-matrices and $ \mathcal{D}_\mu $ stands for the 
fermionic covariant derivative.

\begin{figure}[h]
\begin{turn}{-90}
\psfrag{"Rnvef1.dat"}{$ s = 0.1 $}
\psfrag{"Rnvef3.dat"}{$ s = 350 $}
\psfrag{"Rnvef5.dat"}{$ s = 440 $}
\psfrag{"Rnvef7.dat"}{$ s = 500 $}
\includegraphics[height=15cm,width=10cm]{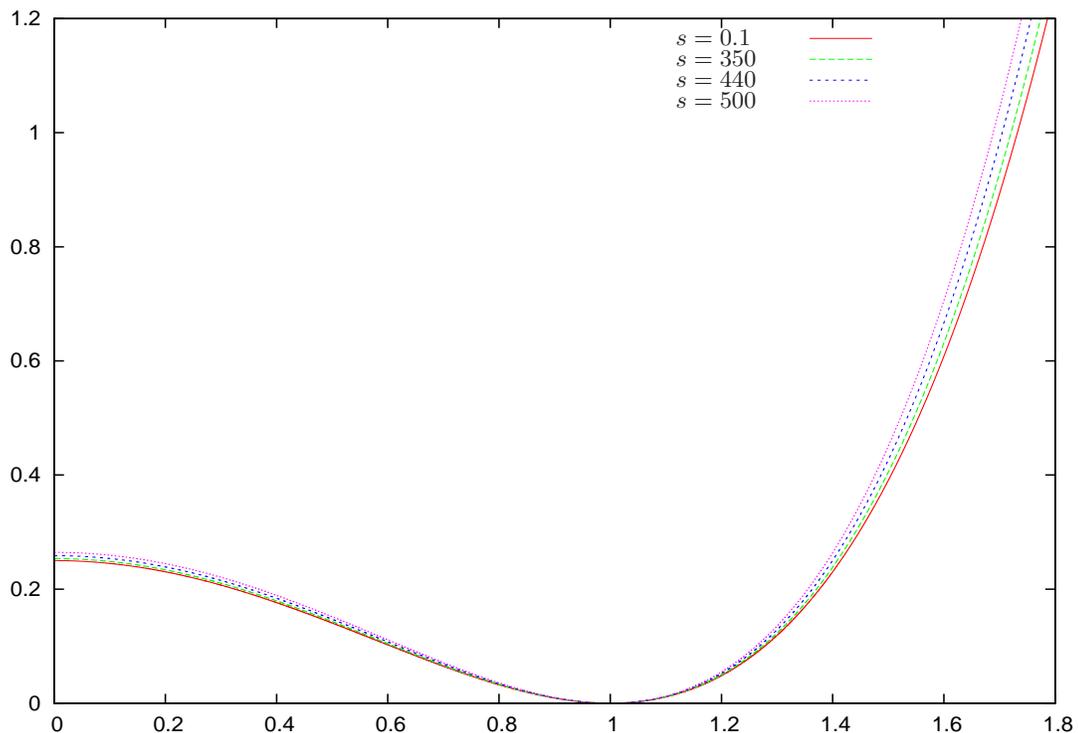}
\end{turn}
\caption{The effective potential $ v_1(u) $ generated from fermions
eq.(\ref{v1fin}) vs. $ u $ for several values of the rescaled Yukawa 
coupling $ s $. The steeper potential corresponds to the largest $ s $. The 
shallower potential corresponds to $ s \to 0 $ and it is the quadratic 
plus quartic potential eq.(\ref{binon}).}
\label{fvef}
\end{figure}

\begin{figure}[h]
\begin{turn}{-90}
\psfrag{"2fnsr1.dat"}{$ s \to 0 $}
\psfrag{"2fnsr7.dat"}{$ s = 500 $}
\psfrag{"2fnsr8.dat"}{$ s = 600 $}
\psfrag{"2fnsr9.dat"}{$ s = 700 $}
\psfrag{"2fnsr10.dat"}{$ s = 800 $}
\includegraphics[height=15cm,width=10cm]{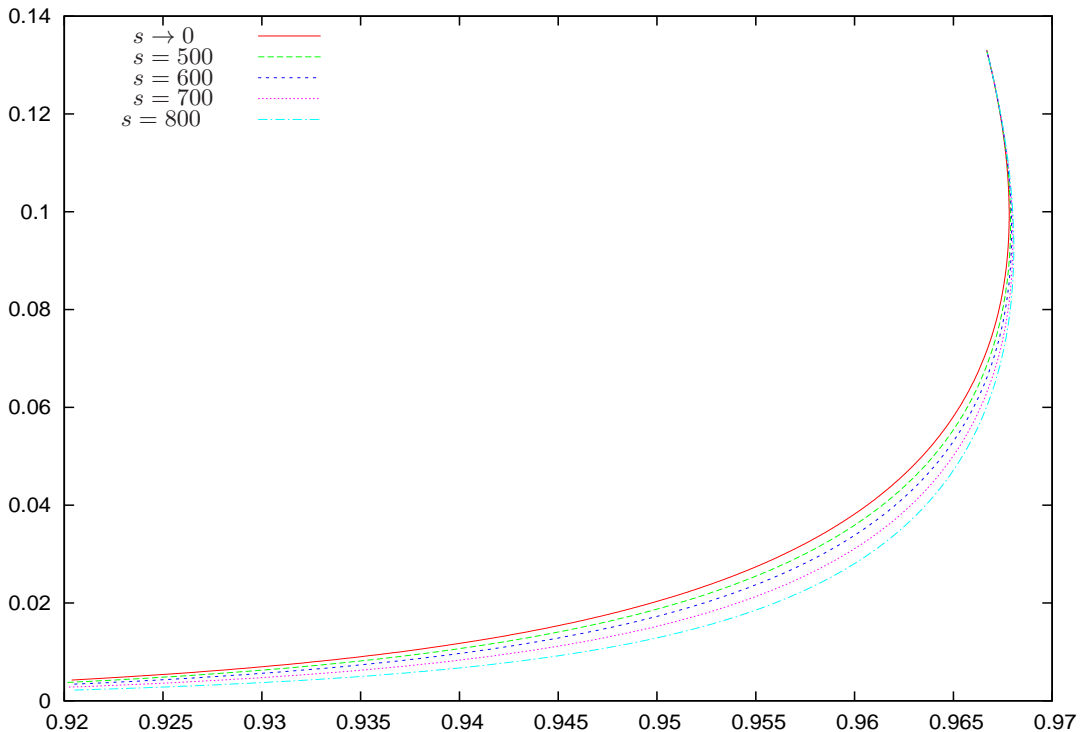}
\end{turn}
\caption{We plot here $ r $ vs. $ n_s $ for the effective potential
obtained from fermions in de Sitter stage eq.(\ref{v1f}) for the physical
value of the parameter $ q $ eq.(\ref{Z2}).
For weak Yukawa coupling $ s \ll 1 $ we recover the $ r=r( n_s) $ curve 
for the quadratic plus quartic potential eq.(\ref{binon}). The  $ r=r( n_s) $ curves
are inside the universal banana region [fig. \ref{banana}] provided  
$ s \leq 850 $, slightly exceeding the bound eq.(\ref{cotaY}).}
\label{nsref}
\end{figure}

For our purposes in this section, the inflationary stage can be approximated by a de 
Sitter space-time (that is, we neglect the slow decrease in time during 
inflation of the Hubble parameter $ H $). In this way, the effective 
potential of fermions can be computed in 
close form with the result \cite{quant,bibl},
\be\label{Vfermi}
  V_f(\varphi) = V_0 + \frac12 \, \mu^2 \, \varphi^2 + 
\frac14 \, \lambda \varphi^4 + H^4\,Q\left(g_Y \,\frac{\varphi}{H}\right)\;,
\ee
where,
\be \label{Vfermi2}
  \begin{split}
    Q(x) &= \frac{x^2}{8\,\pi^2}\left\{(1+x^2)\left[\gamma + 
        \mathrm{Re}\,\psi(1+i\,x)\right] - \zeta(3)\,x^2 \right\}
\quad , \quad x \equiv g_Y \,\dfrac{\varphi}{H} \; , \\
    &= \frac{x^4}{8\,\pi^2} \left[(1+x^2) \sum_{n=1}^{\infty} 
      \frac1{n\,(n^2 + x^2)} - \zeta(3) \right] \\
    &= \frac{x^4}{8\,\pi^2} \sum_{n=1}^{\infty} (-1)^{n+1} \, 
    \left[\zeta(2\,n+1)-\zeta(2\,n+3)\right] x^{2\,n} \;.
  \end{split}
\ee
We included in $ V_f(\varphi) $ the renormalized mass $ \mu^2 $ and 
renormalized coupling 
constant $ \lambda $ which are free and finite parameters. $ \psi(x) $ 
stands for the digamma function, $ \gamma $ for the Euler-Mascheroni 
constant and $ \zeta(x) $ for the Riemann zeta function \cite{PBM}.

Eq.(\ref{Vfermi}) is the energy density for an homogeneous
inflaton field $ \varphi $ coupled to massless fermions through the 
Lagrangian eq.(\ref{Lf}) in a de Sitter space-time. 

The power series of the function $ Q(x) $ has coefficients with
alternating signs, but it can be readily verified that $ Q(x)>0 $ and 
$ Q'(x)>0 $ for $ x>0 $. Moreover, to leading order we have
\be\label{sapr}
Q(x) \buildrel{x \to \infty}\over=  \frac{x^4}{8\,\pi^2}\left[\log x + 
\gamma - \zeta(3) + {\cal O}\left(\frac1{x}\right) \right] \; .
\ee
The constant $ V_0 $ in eq.~(\ref{Vfermi}) must be such that the potential $
V_f(\varphi) $ fulfills eq.~(\ref{minV}) producing a finite number of inflaton
efolds. We consider new inflation and choose $ \mu^2 = -m^2 < 0 $. Hence
$ V_f(\varphi) $ has a double--well shape with the absolute minimum at
$ \varphi=\varphi_{min} $, with $ \varphi_{min} $ a function of the free 
parameters of the potential. 

Expanding $ V_f(\varphi) $ in powers of $ \varphi $ gives
$$
V_f(\varphi) = V_0 - \frac12 \; m^2 \; \varphi^2 + 
\frac14 \; \lambda \; \varphi^4 +\frac1{8 \; \pi^2} \; \left[\zeta(3)
-\zeta(5) \right] \frac{(g_Y \; \varphi)^6}{H^2} +
{\cal O}\left(g_Y^8 \; \varphi^8\right) \; .
$$
where $ \zeta(3)-\zeta(5) = 0.16513 \ldots > 0 $.

In the $ H \to 0 $ limit, eq.(\ref{Vfermi}) becomes
the effective potential for fermions in Minkowski space-time \cite{bibl}.
Recall that 
\be\label{HN}
H = \sqrt{N} \; {\cal H} \; m \quad , \quad m=\frac{M^2}{M_{Pl}}
\ee
where the dimensionless Hubble parameter $ \cal H $ turns out
to be of order one \cite{bibl}. 

\medskip

As in the general description of section \ref{sec2}, eq.(\ref{defg}) we
introduce the dimensionless coupling constant $ g $ as
\begin{equation*}
  g = \frac{M_{Pl}^2}{\varphi_{min}^2} 
\end{equation*}

Besides $ g $ we can now form two other independent and positive 
dimensionless shape parameters, that is
\be\label{sz}
s \equiv g_Y \; \frac{\varphi_{min}}{H} = \frac{g_Y}{\sqrt{g}} \; 
\frac{M_{Pl}}{H}
  \quad, \quad q \equiv \frac{H^2}{m \; \varphi_{min}} =
  \sqrt{g} \; \frac{H^2}{m \; M_{Pl}} = \sqrt{g} \; \frac{H^2}{M^2} 
  \quad, \quad x = s \; u \; .
\ee
We derive the dimensionless potential $ v_1(u) $ from eq.(\ref{Vfermi}) 
using the general transformation equations (\ref{wvx}). We obtain,
\be\label{v1f}
v_1(u) = \frac{g}{M^4} \; V_f(\varphi_{min}\,u) =  c_0 - \frac12 \; u^2 + 
\frac14 \; c_4 \; u^4 + \frac{g}{M^4} \; H^4 \; Q(s \; u)
\ee
The parameters  $ c_0 $ and $ c_4 $ are determined by requiring that $ 
v_1(1)=v_1'(1)=0 $ as in sec. \ref{sec2}. We thus obtain from eq.(\ref{v1f})
\be\label{c04}
c_4 = 1 - q^2 \; s \; Q'(s)  \quad, \quad c_0 = \frac14 + \frac14 \; q^2 
\; s \;  Q'(s) -  q^2 \; Q(s)  
\ee
Inserting  $ c_0 $ and $ c_4 $ into eq.(\ref{v1f}) yields for the 
inflaton potential
\be\label{v1fin}
  \begin{split}
    v_1(u) = \frac{g}{M^4} \; V_f(\varphi_{min}\,u) \;&= 
    \frac12\,(1-u^2) + \frac14 \left[1-q^2\,s\,Q'(s)\right]\, (u^4-1) 
    + q^2\left[Q(s\,u) - Q(s)\right] \\ &= \frac12\,(1-u^2) + 
    \frac14(1-b)\, (u^4-1) + b\,F(u,s) \;,
  \end{split}
\ee
where
\be
b \equiv q^2 \; s \; Q'(s) \ge 0\; , \quad  F(u,s) \equiv \frac{Q(s\,u) - 
Q(s)}{s \; Q'(s)}
\ee
Notice that $ v_1(u) $ reduces to the quartic double--well potential 
$ \frac14 (1-u^2)^2 $ when $ s\to 0 $ at fixed
$ q $ (that is, when $ g_Y\to 0 $) as well as when $ b\to0 $ at fixed $ s $.
This last limit means $ g_Y\to 0 $ with $ g_Y \; M_{Pl} / H $ fixed.

\medskip

Only the interval $ 0<u<1 $ is relevant for the inflaton evolution.
For any $ u $ in this interval, $ F(u,s) $ is negative definite and is 
monotonically decreasing as a function of $ s $. In particular,
\begin{equation*}
  F(u,s) \buildrel{s \to 0}\over= \frac16 \; (u^6-1) \quad , \quad 
  F(u,s) \buildrel{s \to \infty}\over= \frac14(u^4-1) +
  {\cal O}\left(\frac1{\log s}\right)\;,\qquad 0<u<1~~\textrm{fixed} 
\end{equation*}
Hence for $ s\to0 $ at fixed $b$ we obtain again the sixth--order 
double--well potential of eq.~(\ref{v1s})
\begin{equation*}
  v_1(u) \longrightarrow 
\frac1{12} \; (1-u^2)^2 \; (3+b+2\,b \; u^2) \quad , \quad
  s\to0~\textrm{at fixed}~b \;,
\end{equation*}
while $ b $ cancels out for large $ s $ and  we get back the quartic 
double--well potential
\begin{equation*}
  v_1(u) \longrightarrow  \frac14 \; (1-u^2)^2 \;,\quad
  s\to\infty~\textrm{at fixed}~b\;.
\end{equation*}
The terms containing $ Q $ in the effective potential eqs.(\ref{Vfermi}) 
and (\ref{v1fin}) represent the one--loop quantum contributions. They vanish 
when $ b=0 $ while for $ b>0 $ they should be sizably smaller than the
tree level contribution, otherwise all higher loops effects must also be
taken into account. In particular, the quartic term in 
eq.~(\ref{v1fin}) must have a non-negative coefficient, that is $ 0 < b\le 1 $ 
as in sec. \ref{sextico}.

We see from eq.(\ref{v1fin}) that the one-loop ($ Q $) pieces are of the 
order $ q^2 \; Q(s \; u) $ compared with the tree-level pieces. 
We can compute $ q $ using eqs.(\ref{gy}), (\ref{HN}) and (\ref{sz}) with 
the result
\be\label{Z2}
q = \sqrt{\frac{N \; y}8} \; 
\left(\frac{{\cal H} \; M}{M_{Pl}}\right)^2 \simeq 0.854 \; 10^{-5} \ll 1 \; ,
\ee
where $ y \simeq 1.3 $ and $ {\cal H} \simeq  0.5 $ \cite{bibl}.

Hence, the one-loop pieces are negligible unless
$ s \gg 1 $. We can therefore use the asymptotic behavior eq.(\ref{sapr})
to estimate  $ Q(s \; u) $ for large $ s $. In order the one-loop part to be 
smaller or of the order of the tree level piece we must impose in the 
strong coupling regime $ s \gg 1 $,
\be\label{cotaY}
\frac1{8 \; \pi^2} \; s^4 \; q^2 \; \ln s \lesssim 1 \quad \Rightarrow \quad s 
\lesssim 1020 \; [\ln 1020]^{-1/4} \simeq 616 \; .
\ee
The one-loop potential  eq.(\ref{v1fin}) is therefore reliable for $ s \lesssim 600 $.
For larger values of $ s $ the one-loop piece is larger than the tree level part and hence
all higher order loops should be included too.

For $ s \sim 1 $  
we recover the quadratic plus quartic potential eq.(\ref{binon})
since the terms in $ s^4 \; q^2 $ are negligible in eq.(\ref{v1fin})
and $ n_s $ and $ r $ are thus given by eqs.(\ref{binsr}).
We find from eqs.(\ref{v1fin}) and (\ref{Z2}) in the case $ s \sim 1 $,
$$
v_1(u)=  \frac14 - \frac12 \; u^2 + \frac14 \; u^4 + 
{\cal O}\left(\frac{H^2}{M_{Pl}^2}\right)
$$
That is, the terms beyond $ u^4 $ in the effective potential from the 
fermions are of the same order of magnitude as the loop corrections to 
inflation \cite{bibl,quan} and can be neglected since 
$ (H/M_{Pl})^2 \sim 10^{-9} $.

\medskip

We display in fig. \ref{nsref} $ r $ vs. $ n_s $ for various values of 
the Yukawa coupling $ s $.
Therefore, the banana region $ \cal B $ in the $ (n_s,r) $ plane 
for the effective potential eq.(\ref{v1fin}) is the region limited
by the curves for the potential for $ s \leq 500 $ and for  
$ s \to 0 $ as displayed in fig. \ref{nsref}.
Notice that the lower border of the region $ \cal B $ for the
effective potential eq.(\ref{v1fin}) is well {\bf above} the 
lower border of the universal $ \cal B $ region displayed in fig. \ref{banana}.

In summary, the $ r=r(n_s) $ curves for the dynamically generated inflaton
potential eq.(\ref{v1fin}) are {\bf inside} the universal banana region $ \cal B $ for 
all values
of the Yukawa coupling $ g_Y $ that keep the result for this one-loop potential reliable.
Namely, the one-loop piece is smaller or of the order of the tree level part.

\section{The Universal banana region $ \cal B $}

\begin{figure}[h]
\psfrag{"nsrN4.dat"}{Quadratic plus quartic potential}
\psfrag{"nsrinf2.dat"}{Quadratic plus $ u^{\infty} $ potential}
\includegraphics[height=12cm,width=14cm]{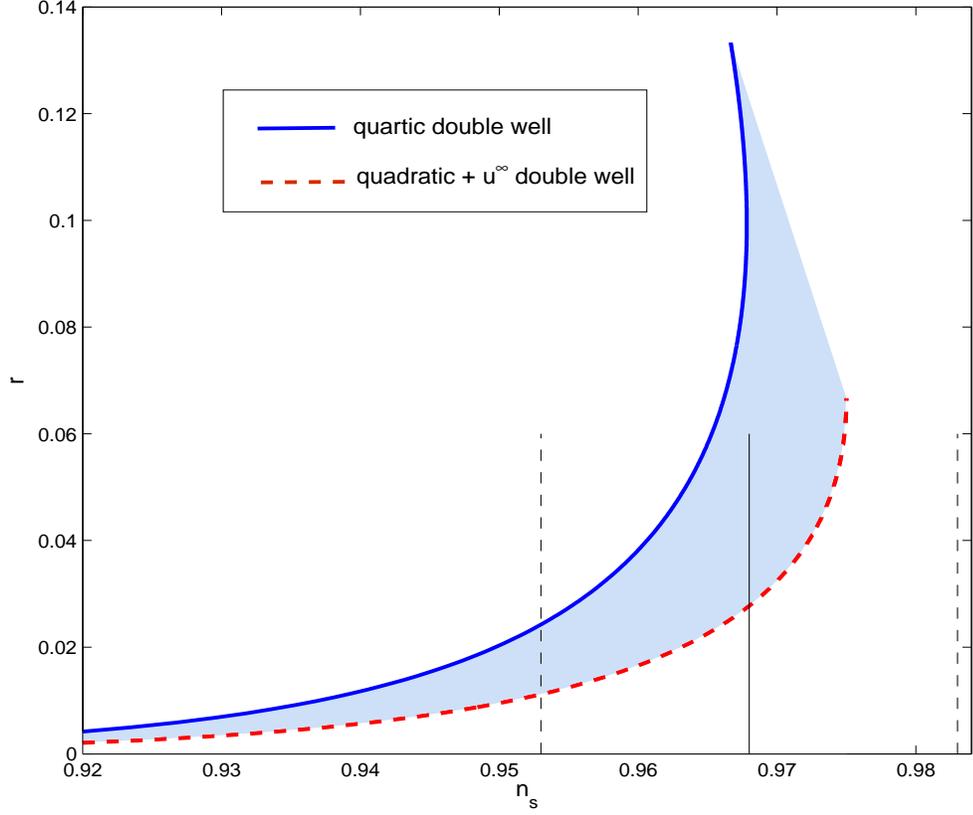}
\caption{We plot here the borders of the universal 
banana region $ \cal B $
in the $ (n_s, r) $-plane 
setting $ N = 60 $. The curves are computed with the quadratic 
plus quartic potential eq.(\ref{binon}) and with the $ n = \infty $ limit 
of the quadratic plus $ u^{2 \, n} $ potential eq.(\ref{wn}) (or the 
$ b = \infty $ limit of the quadratic plus exponential potential
eq.(\ref{expo}), which gives identical results) as given by 
eqs.(\ref{ginf2})-(\ref{nsrinf}) and eqs.(\ref{ginf}) and 
(\ref{rnsinf}). Notice that the lower part of the right border of $ \cal B $,
$ 0 < r < 4/N = 0.06666\ldots $ corresponds to the limit
$ n = \infty $ at fixed $ u $ eq.(\ref{nsrinf}). 
The upper part $ 4/N < r < 8/N $ of the right border
of $ \cal B $ is not displayed here. We display in the vertical full line
the LCDM+r value $ n_s = 0.968 \pm 0.015 $ using WMAP5+BAO+SN data.
The broken vertical lines delimit the $ \pm 1 \, \sigma$ region.}
\label{banana}
\end{figure}

In summary, we find that all $ r=r(n_s) $ curves for double--well inflaton 
potentials in the Ginsburg-Landau spirit fall {\bf inside} the {\bf universal}
banana region $ \cal B $ depicted in fig. \ref{banana} for new 
inflation.  Namely,
\begin{itemize}
\item{The fourth degree double--well potentials containing a cubic term
studied in ref. \cite{mcmc}:
\be\label{cuart}
v_1(u) = \frac14 +\frac{\beta}6  - \frac12 \; u^2 -\frac23 \; \beta \; u^3
+ \frac14 (1+2 \, \beta) \; u^4 \; ,
\ee
where $ \beta \geq 0 $ is the asymmetry parameter. This potential reduces to eq.(\ref{binon}) 
for $ \beta = 0 $.}
\item{The quadratic plus sixth-order potential eq.(\ref{v1s}).}
\item{The even polynomial potentials with arbitrarily
higher--order degrees and positive coefficients
 (sec. \ref{pares}).}
\item{The quadratic plus exponential potential (sec. \ref{potex}).}
\item{The inflaton potential dynamically generated from fermions (sec. \ref{potferm}).}
\end{itemize}
Potentials in the Ginsburg-Landau spirit have usually coefficients of order
one when written in dimensionless variables. This is the case of the inflaton potentials 
$ v_1(u) $. In that case, we found that all $ r=r(n_s) $ curves for 
double--well potentials fall {\bf inside} the universal banana region $ \cal B $ 
depicted in fig. \ref{banana}. Moreover, for even double--well potentials with arbitrarily large
positive coefficients, their $ r=r(n_s) $ curves 
lie inside the universal banana region $ \cal B $ [fig. \ref{banana}].

\medskip

The study of the dynamically-generated inflaton potential in sec. \ref{potferm} leads to
analogous conclusions. This one-loop inflaton potential is reliable as long as 
the one-loop piece is smaller or of the same order than the tree level part.
In such regime all the curves  $ r = r(n_s) $ produced by this fermion-generated potential
lie {\bf inside} the universal banana region $ \cal B $ 

More generally, we see from eq.(\ref{rygran}) that $ r \ll 1 $ is generally linked
to a large coupling $ y \gg 1 $. However, this strong coupling regime corresponds to 
$ n_s $ values well below the current best observed value $ n_s = 0.964 $, and 
is therefore excluded by observations.

\medskip

The lower border of the universal region $ \cal B $ corresponds to the limit
binomial potential eq.(\ref{limv}) 
$$
v_1(u) = \frac12 \; (1-u^2) \quad {\rm for} \; u<1 \quad , \quad 
v_1(u) = +\infty\quad {\rm for} \; u>1 \; .
$$
and is described parametrically by eq.(\ref{nsrinf}).
We obtain such potential and such parametrization of $ r=r(n_s) $ both
as the $ n \to \infty $ limit of the quadratic plus $ u^{2 \, n} $ potential
in sec. \ref{pot2n} as well as the $ b \to \infty $ limit of the 
$ e^{2 \, b \; g \; \phi^2} $ potential in sec \ref{potex}. 

The upper-right border of the universal banana-shaped region $ \cal B $ is not given
by eqs.(\ref{ginf}) and (\ref{rnsinf}) corresponding to the double limit
$ n \to \infty $ and $ u \to 1 $
(or, alternatively $ b \to \infty $ and $ u \to 1 $). This follows from 
the fact that some potentials of order $ 100 $ yield $ r=r(n_s) $ curves
above the limiting curves for the quadratic plus $ u^{100} $ potential
as depicted in fig. \ref{fig:ban100}.

The upper-left border of the universal region $ \cal B $ depicted in 
fig. \ref{banana} is given by the fourth order double--well potential
eq.(\ref{binon}) and it is described parametrically by eq.(\ref{binsr}).

The lower border of the universal region $ \cal B $ is particularly relevant
since it gives a {\bf lower bound} for $ r $ for each observationally
allowed value of $ n_s $. For example, the best $ n_s $ value $ n_s = 0.964 $
implies from fig. \ref{banana} that $ r > 0.021 $.

The upper border of the universal region $ \cal B $ tells us the upper bound
$ r < 0.053 $ for $ n_s = 0.964 $.

Therefore, we have within the large class of potentials inside the region $ \cal B $
\be\label{cotaB}
 0.021 < r < 0.053 \quad {\rm for}\quad  n_s = 0.964 \; .
\ee
Notice that these bounds on $ r $ are compatible with the experiments \cite{WMAP5}.

\medskip

The Ginsburg-Landau criterion applied to the inflaton potential 
eq.(\ref{vpol2}) leads to values of $ (n_s, \; r) $ inside the 
universal banana-shaped region $ \cal B $.
In principle, within the Ginsburg-Landau
approach, one should choose potentials $ v_1(u) $ with {\bf small} 
coefficients $ |c_k| \gg 1 $ in eq.(\ref{v1gen}). However, as shown 
in sec. III, this restriction does not apply for double--well potentials 
of even degree: for this class of potentials {\bf large} and positive
coefficients $ c_{2 \, k} \gg 1 $ also provide $ (n_s, \; r) $  
{\bf within} the universal banana region $ \cal B $. Moreover,
the double--well quartic potential eq.(\ref{cuart}) also provides 
$ (n_s, \; r) $ inside the banana region $ \cal B $
for arbitrarily large coefficient $ \beta $ \cite{mcmc}.

\medskip

On the contrary, double--well potentials of degree larger than four
with large negative coefficients are outside the Ginsburg-Landau class and
produce $ (n_s, \; r) $ outside the region $ \cal B $.
In particular, it is possible to produce in this way $ n_s $ values
compatible with the data together with arbitrarily small
values for $ r $ by choosing large enough negative coefficients $ c_k $.

\medskip

We find that the Ginsburg-Landau class of potentials is 
{\bf physically well
motivated} and therefore that the banana-shaped region $ \cal B $
is a natural region to expect to observe $ (n_s, \; r) $.
Namely, taking into account the present data for $ n_s $ we expect that
$ r $ will be observed in the interval eq.(\ref{cotaB}).
Anyhow, the fourth order double--well potential eq.(\ref{bini}) 
provides and excellent fit to the present
CMB/LSS data and yields as most probable values: $ n_s \simeq 0.964 ,\;
r\simeq 0.051 $.

\medskip

The  physical framework provided by the Ginsburg-Landau effective theory 
of inflation allows to take high benefit of the data (of the present
data and the forthcoming ones). The lower bounds 
(and most probable value) we infer for $ r $ are the best way to support 
the searching for CMB polarisation and the future missions on it.

\end{document}